\documentclass[a4paper,12pt, epsfig]{article}
\def\letter{0}\def\pr{0}
\pdfoutput=1 
\usepackage{epsfig}
\usepackage{epstopdf}
\usepackage{graphicx}
\usepackage{ifthen}
\usepackage{ulem}

\usepackage{amsmath}
\usepackage[psamsfonts]{amssymb}
\usepackage{euscript}

\usepackage{latexsym}
\usepackage[arrow,matrix,curve]{xy}

\usepackage[hypertexnames=false]{hyperref}
\hypersetup{
    colorlinks=true,
    linkcolor=blue,
    filecolor=magenta,   
    citecolor=black,   
    urlcolor=cyan,
    }

\newskip\humongous \humongous=0pt plus 1000pt minus 1000pt

\newif\ifdtup

\def\,{\hspace{-.1cm}}
\def\hsp{,\hspace{.7cm}}

\def\fc#1#2 {\frac{n}{q}#1\frac{n}{q}#2}

\def\kt{\kappa}
\def\kt{\mathfrak{K}}
\def\ks{|\kt\rangle}

\def\hp{H^{\prime(t)}}
\def\rv{{\rm{vac}}}

\newcommand{\vac}{\ensuremath{|0\rangle}}

\renewcommand{\sin}{\textrm{sin}}

\renewcommand{\sinh}{\textrm{sinh}}
\renewcommand{\cosh}{\textrm{cosh}}
\renewcommand{\tanh}{\textrm{tanh}}
\newcommand{\sech}{\textrm{sech}}
\newcommand{\csch}{\textrm{csch}}

\def\exp#1{\hbox{\rm exp}\left[#1\right]}

\def\sign#1{{\rm sign}\left(#1\right)}

\renewcommand{\theequation}{\arabic{section}.\arabic{equation}}
\renewcommand{\(}{\begin{equation}}
\renewcommand{\)}{end{equation} \vspace{-.05in}\linebreak}

\newcounter{saveeqn}
\newcounter{savealpheqn}

\newcommand{\alpheqn}{\setcounter{saveeqn}{\value{equation}}%
  \stepcounter{saveeqn}\setcounter{equation}{0}%
  \renewcommand{\theequation}{\mbox{\arabic{section}.\arabic{saveeqn}
\alph{equation}}}
  \renewcommand{\)}{\end{equation}}}
\def\part#1{\frac{\partial}{\partial{#1}}}%
\def\group#1{\refstepcounter{equation}\setcounter{saveeqn}
 {\value{equation}}%
  \label{#1}\setcounter{equation}{0}%
\renewcommand{\theequation}{\mbox{\arabic{section}.\arabic{saveeqn}
\alph{equation}}}
  \renewcommand{\)}{\end{equation}}}
\newcommand{\reseteqn}{\setcounter{equation}{\value{saveeqn}}%
  \renewcommand{\theequation}{\arabic{section}.\arabic{equation}}%
  \renewcommand{\)}{\end{equation}}}

\newcommand{\aalpheqn}{\setcounter{saveeqn}{\value{equation}}%
  \stepcounter{saveeqn}\setcounter{equation}{0}%
  \renewcommand{\theequation}{\mbox{
        \Alph{subsection}.\arabic{saveeqn}\alph{equation}}}
   \renewcommand{\)}{\end{equation}}}
\newcommand{\areseteqn}{\setcounter{equation}{\value{saveeqn}}%
  \renewcommand{\theequation}{\Alph{subsection}.\arabic{equation}}%
  \renewcommand{\)}{\end{equation}}}

\renewcommand{\thefootnote}{\alph{footnote}}
\renewcommand{\(}{\begin{equation}}
\renewcommand{\)}{\end{equation}}
\newcommand{\ba}{\begin{eqnarray}}
\newcommand{\ea}{\end{eqnarray}}

\renewcommand{\b}{\beta}

\renewcommand{\sl}{{\sqrt{\lambda}}}

\newcommand{\cbp}{\mathop{\vtop{\ialign{##\crcr
   $\hfil\displaystyle{}\hfil$\crcr\noalign{\kern-13pt\nointerlineskip}
   \BIG{)}\hskip0pt\crcr\noalign{\kern3pt}}}}}
\newcommand{\pa}{\mathop{\vtop{\ialign{##\crcr

$\hfil\displaystyle{\oplus}\hfil$\crcr\noalign{\kern+1pt\nointerlineskip
}
   \hspace{.08in}$^{\alpha=0}$\hskip6pt\crcr\noalign{\kern3pt}}}}}
\renewcommand{\hsp}{,\hspace{.3in}}
\newcommand{\p}{^\prime}
\newcommand{\pp}{^{\prime\prime}}

\catcode`\@=11
\def\vereq#1#2{\lower3pt\vbox{\baselineskip1.5pt \lineskip1.5pt
\ialign{$\m@th#1\hfill##\hfil$\crcr#2\crcr\sim\crcr}}}
\catcode`\@=12

\renewcommand{\(}{\begin{equation}}
\renewcommand{\)}{\end{equation}}

\def\pin#1{\int \frac{d#1}{2\pi}}
\def\ppin#1{\int\hspace{-17pt}\sum \frac{d#1}{2\pi}}
\def\ppink#1{\int\hspace{-17pt}\sum\frac{d^{#1}k}{(2\pi)^{#1}}}
\def\ppinkp#1{\int\hspace{-17pt}\sum\frac{d^{#1}k\p}{(2\pi)^{#1}}}
\def\dint{\int\hspace{-12pt}\sum }
\def\pink#1{\int \frac{d^{#1}k}{(2\pi)^{#1}}}

\def\pinkp#1{\int \frac{d^{#1}k\p}{(2\pi)^{#1}}}

\def\Bd#1{B^\ddag_{k_{#1}}}
\def\Btd#1{B^{(t)\ddag}_{k_{#1}}}
\def\Bt#1{B^{(t)}_{k_{#1}}}
\def\Bdp#1{B^\ddag_{k\p_{#1}}}

\def\cc{\mathcal{C}}
\def\df{\mathcal{D}_{f}}
\def\dft{\mathcal{D}_{f}^{(t)}}
\def\dfd{\mathcal{D}_{f}^{(t)\dag}}

\def\I{\mathcal{I}}

\def\os{\omega_S}

\def\gx{(\gamma(x-vt))}

\def\blu#1{\textcolor{blue}{Jarah: #1}}
\def\red#1{\textcolor{red}{Kehinde: #1}}

\newcommand{\beas}{\begin{eqnarray*}}
\newcommand{\eeas}{\end{eqnarray*}}

\newcommand{\bquo}{\begin{quote}}
\newcommand{\enqu}{\end{quote}}

\def\lim#1{\stackrel{\rm{lim}}{{}_{#1}}}

\newcommand{\R}{{\mathbb R}}

\newcommand{\g}{\mathfrak g}

\def\ch{{\mathcal{H}}}
\def\co{{\mathcal{O}}}

\def\ok#1{\omega_{k_{#1}}}
\def\okp#1{\omega_{k\p_{#1}}}
\def\okt#1{\omega_{\kt_{#1}}}
\def\V#1{V^{(#1)}(\sqrt{\lambda}f(x))}
\def\Vg#1{V^{(#1)}(\sqrt{\lambda}f(\gamma(x-vt))}
\def\v#1{V^{(#1)}[f(x),x]}
\def\ck{\csch\left(\frac{\pi k}{2\b}\right)}

\def\mb{\mathcal{B}}
\def\mc{\mathcal{C}}
\def\md{\mathcal{D}}
\def\me{\mathcal{E}}

\newcommand{\beq}{\begin{equation}}
\newcommand{\eeq}{\end{equation}}
\newcommand{\bea}{\begin{eqnarray}}
\newcommand{\eea}{\end{eqnarray}}

\newskip\humongous \humongous=0pt plus 1000pt minus 1000pt

\newif\ifdtup

\catcode`\@=11

\ifthenelse{\equal{\letter}{0}}{ 

\@addtoreset{equation}{section}
\def\theequation{\arabic{section}.\arabic{equation}}

\def\@normalsize{\@setsize\normalsize{15pt}\xiipt\@xiipt
\abovedisplayskip 14pt plus3pt minus3pt%
\belowdisplayskip \abovedisplayskip
\abovedisplayshortskip \z@ plus3pt%
\belowdisplayshortskip 7pt plus3.5pt minus0pt}

\def\small{\@setsize\small{13.6pt}\xipt\@xipt
\abovedisplayskip 13pt plus3pt minus3pt%
\belowdisplayskip \abovedisplayskip
\abovedisplayshortskip \z@ plus3pt%
\belowdisplayshortskip 7pt plus3.5pt minus0pt
\def\@listi{\parsep 4.5pt plus 2pt minus 1pt
      \itemsep \parsep
      \topsep 9pt plus 3pt minus 3pt}}

\relax

\def\section{\@startsection{section}{1}{\z@}{3.5ex plus 1ex minus  .2ex}{2.3ex plus .2ex}{\large\bf}}

\def\thesection{\arabic{section}}
\def\thesubsection{\arabic{section}.\arabic{subsection}}

\def\appendix{\setcounter{section}{0}
 \def\thesection{Appendix \Alph{section}}
 \def\thesubsection{\Alph{section}.\arabic{subsection}}
 \def\theequation{\Alph{section}.\arabic{equation}}}
\renewcommand{\theequation}{\arabic{section}.\arabic{equation}}

}{
\renewcommand{\theequation}{\arabic{equation}}

} 

\begin{document}
\def\thefootnote{\fnsymbol{footnote}}
\def\thetitle{Perturbative Approach to Time-Dependent Quantum Solitons}
\def\autone{Kehinde Ogundipe}
\def\auttwo{Jarah Evslin}
\def\affc{Institute of Physics, Polish Academy of Sciences, Aleja Lotników 32/46, 02-668 Warsaw, Poland}
\def\affb{University of the Chinese Academy of Sciences, YuQuanLu 19A, Beijing 100049, China}
\def\affa{Institute of Modern Physics, NanChangLu 509, Lanzhou 730000, China}


\ifthenelse{\equal{\pr}{1}}{
\title{\thetitle}
\author{\autone}
\author{\auttwo}
\author{\autthree}
\affiliation {\affa}
\affiliation {\affb}

}{

\begin{center}
{\large {\bf \thetitle}}

\bigskip

\bigskip


{\large \noindent  \autone
\footnote{kehindeoogundipe@gmail.com} 
\ and \auttwo
\footnote{jarah@impcas.ac.cn}
}


\vskip.7cm

1) \affa\\
2) \affb\\

\end{center}

}

\begin{abstract}
\noindent
Recently we have introduced a lightweight, perturbative approach to quantum solitons.  
Thus far, our approach has been largely limited to configurations consisting of a single soliton plus a finite number of mesons, whose classical limit is an isolated stationary or rigidly moving soliton.  In this paper, with an eye to soliton collisions and oscillons, we generalize this approach to quantum states whose classical limits are genuinely time-dependent.  More precisely, we use a unitary operator, inspired by the coherent state approach to solitons, to factor out the nonperturbative part of the state, which includes the classical motion.  The solution for the quantum state and its evolution is then reduced to an entirely perturbative problem.

\end{abstract}

%
\setcounter{footnote}{0}
\renewcommand{\thefootnote}{\arabic{footnote}}

\ifthenelse{\equal{\pr}{1}}
{
\maketitle
}{}

\section{Introduction}
\subsection{Motivation}
The present work is motivated by three questions.  The first concerns oscillons.  Oscillons are time-dependent classical solutions whose existence depends on the sign of a leading nonlinearity \cite{sk}.  Therefore, in a sense, they are present in half of all theories and these include many of phenomenological interest.   Oscillons are generally \cite{osc1,osc2,osc3} created by any violent event, such as a first order phase transition or soliton collision.  They are unstable, yet in general they are the longest lived unstable excitations.  As a result, at intermediate timescales after a violent event they dominate the dynamics.  

However, it is widely believed that the situation in the quantum theory is very different.  Refs.~\cite{hertzberg,tanmay} have argued that in the quantum theory oscillons experience a rapid decay into pairs of elementary quanta.  Therefore, it is believed that once quantum corrections are considered, oscillons do not dominate the dynamics at any time.  In Ref.~\cite{noiosc}, it was argued that one-loop corrections to the states might radically alter these conclusions.  Yet such corrections have never been calculated.  We aim to present a formalism which will allow the calculation of such corrections.

Second, it has long been known \cite{sug,csw,frac91} that kink-antikink collisions lead to an interesting fractal pattern of escape windows.  Needless to say, fractal patterns can be affected qualitatively by even arbitrarily small perturbations \cite{pert0,pert1,pert2,pert3}.  Thus various authors have wondered throughout the ages just what effect the leading quantum corrections have on this resonance structure, although the first serious study of this question appeared only quite recently~\cite{qpert23}.

These first two questions are rather straightforward applications.  The third is more speculative.  The Unruh effect in many ways resembles the two-particle decay claimed for oscillons in Refs.~\cite{hertzberg,tanmay}.  It is therefore interesting, in our opinion to calculate the leading quantum corrections to Rindler space states and to see how they affect the radiation spectrum.  Indeed, an intriguing paper \cite{akh21} has recently suggested that the usual choice of states, on the future wedge, leads to a secular evolution, similarly to the choice of oscillon states in the above works, suggesting that such quantum corrections indeed arise automatically in the natural course of evolution. 

\subsection{Goals}

To attack these problems, one first needs to choose a formalism for treating quantum states corresponding to classically nontrivial configurations.  Several such formalisms are available.  At the linear level there are powerful spectral methods \cite{gw22} and also the classical-quantum correspondence of Refs.~\cite{cq1,cq2}.  However, we are interested in not only linear but also the higher order nonlinear effects.  The usual tool of choice in this regime is the collective coordinate method of Ref.~\cite{gs74}.  However, after separating the collective coordinates, one needs a complicated canonical transformation to return to canonical coordinates, leading to an infinite number of terms in the Hamiltonian.  An additional infinite number of terms is required \cite{gj76} at the nonlinear level so that this transformation will leave the path integral measure invariant.  In the end, all of these terms lead to a formalism which is powerful but also cumbersome.

We therefore plan to address all of these questions using the linearized soliton perturbation theory introduced in Refs.~\cite{mekink,me2loop}.  This is a variant of the strategy in Ref.~\cite{cahill76} which avoids a notorious problem \cite{rebhan} arising from separately regularizing the vacuum and nontrivial sectors by regularizing only once, and then using a unitary transformation that is equivalent to the coherent state construction of solitons \cite{taylor78}, with the necessary \cite{cocorr23} corrections, to relate the sectors.

So far this formalism has largely been applied to states that are obtained by quantizing time-independent solutions of the classical equations of motion.  That is not to say that dynamics has not been studied, indeed kink-meson scattering amplitudes have been calculated exhaustively at the leading order \cite{memult,mestokes}, and also the elastic kink-meson scattering amplitude has recently been calculated \cite{elastico}.  However, the soliton is heavier than the perturbative degrees of freedom by a factor of the inverse coupling $1/\lambda$, and so the effect of these perturbative processes on the the soliton  is suppressed by the coupling $\lambda$.  As $\lambda\hbar$ is dimensionless, the perturbative degrees of freedom vanish in the classical limit $\hbar\rightarrow 0$.  Thus classically there was no dynamics.

For the three motivating questions, we are interested in a different regime.  We are interested in classical solutions that are themselves time-dependent, corresponding to a nonperturbative time dependence in the quantum field theory.

How can we treat a nonperturbative time-dependence using perturbation theory?  We do it in the same way as we treat a nonperturbative soliton in perturbation theory:  We use a nonperturbative unitary transformation to separate out the nonperturbative part.  We refer to the transformed states and operators as the comoving frame.  The introduction and study of this frame are the main results of this paper.

\subsection{Summary}

There is one case in which linearized soliton perturbation theory has already been applied to a classically time-dependent soliton.  This is the case of the moving kink, treated in Ref.~\cite{memove} and used to calculate form factors in Refs.~\cite{meff,hengyuan}.  This was possible, without the introduction of the comoving frame, because it is related by a Lorentz transformation to a solution with no time-dependence.   Therefore our strategy will be to use this example to learn how to construct the comoving frame in general, for solutions that may not be kinks at all.  

We begin in Sec.~\ref{classsez} with a quick review of the classical kink.  We review the quantization of the stationary kink in Sec.~\ref{quantsez}.  Then we quantize the moving kink in Sec.~\ref{movesez}.  We see the obstructions to a perturbative approach.  Therefore, in Sec.~\ref{cosez} we introduce the comoving frame, in the case of the moving kink.  Once we have understood the comoving frame in this example, the generalization to an arbitrary time-dependent solution of the classical equations of motion is clear, and so we present our general construction in Sec.~\ref{gensez}.  The reader uninterested in the intuitive derivation via the rigidly moving kink can skip straight to Sec.~\ref{gensez} to see our results.

\section{Classical Kink} \label{classsez}

We believe that the results of this paper can be readily generalized to more phenomenologically interesting theories.  However, we will work in a 1+1 dimensional theory of a scalar $\phi(x)$ and its conjugate $\pi(x)$, described by the Hamiltonian 
\beq
H=\int d x: \mathcal{H}(x):_a, \quad \mathcal{H}(x)=\frac{\pi^2(x)}{2}+\frac{\left(\partial_x \phi(x)\right)^2}{2}+\frac{V(\sqrt{\lambda} \phi(x))}{\lambda}. \label{heq}
\eeq
Here $\lambda$ is the coupling constant, $V$ is a degenerate potential and we have included the usual normal ordering $::_a$ which acts trivially in the classical theory, but eliminates all ultraviolet divergences in the quantum theory, which we turn to in Sec.~\ref{quantsez}.  If one wishes to add more dimensions or fermions, then counterterms will be necessary to treat the corresponding divergences.  

The classical equations of motion are
\beq
\dot\phi(x,t)=\frac{\delta H}{\delta \pi(x)}=\pi(x,t)
\hsp
\dot\pi(x,t)=-\frac{\delta H}{\delta \phi(x)}=\partial_x^2\phi(x,t)-\frac{V^{(1)}(\sl\phi(x,t))}{\sl}
\eeq
where we have defined
\beq
V^{(n)}(\sqrt{\lambda} \phi(x))=\frac{\partial^n V(\sqrt{\lambda} \phi(x))}{\partial (\sqrt{\lambda} \phi(x))^n}.
\eeq
Putting these together yields the classical equation of motion
\beq
\ddot\phi(x,t)-\partial_x^2\phi(x,t)+\frac{V^{(1)}(\sl\phi(x,t))}{\sl} =0.\label{eom}
\eeq

\subsection{Stationary Classical Kink}

Consider a static kink solution
\beq
\phi(x,t)=f(x)\hsp f\pp(x)=\frac{\V1}{\sl}. \label{bps}
\eeq
The corresponding classical energy is
\beq
Q_0=\int dx \left[ \frac{f^{\prime 2}(x)}{2}+\frac{V(\sqrt{\lambda} f(x))}{\lambda}
\right]. \label{q0}
\eeq

Let us multiply the rightmost expression in Eq.~(\ref{bps}) by $f\p(x)$
\beq
\frac{1}{2}\partial_x f^{\p 2}(x)=f\pp(x)f\p(x)=\frac{\V1 f\p(x)}{\sl}=\frac{\partial_x V(\sl f(x))}{\lambda}
\eeq
and integrate, yielding
\beq
\frac{f^{\prime 2}(x)}{2}=\frac{V(\sl f(x))}{\lambda}+C
\eeq
where $C$ is a constant of integration.

Let
\beq
V(\pm\infty)=0
\eeq
then
\beq
C=0\hsp \frac{f^{\prime 2}(x)}{2}=\frac{V(\sl f(x))}{\lambda}.
\eeq
So we learn that the two terms in (\ref{q0}) are equal
\beq
\frac{Q_0}{2}=\int dx  \frac{f^{\prime 2}(x)}{2} =\int dx  \frac{V(\sqrt{\lambda} f(x))}{\lambda}.
\eeq

Consider a small perturbation
\beq
\phi(x,t)=f(x)+\g(x) e^{\pm i\omega t}.
\eeq
The linearized equation of motion is the Sturm-Liouville equation
\bea
0&=&\ddot\phi(x,t)-\partial_x^2\phi(x,t)+\frac{V^{(1)}(\sl\phi(x,t))}{\sl} \label{sl} \\
&=&\left[-\omega^2 \g(x)-\g\pp(x)+\V2 \g(x)
\right]e^{\pm i\omega t}.\nonumber
\eea
There are three kinds of solutions, classified by their frequencies $\omega$.  There is always a unique, real zero mode $\g_B(x)$ with $\omega_B=0$.  For every real $k$ there is a continuum normal mode $\g_k(x)$ with $\ok{}=\sqrt{m^2+k^2}$.  Finally there may be discrete, real shape modes $\g_S(x)$ with $0<\omega_S<m$.  We will define the meson mass $m$ momentarily.  We will fix the normalizations of the normal modes using the completeness relation
\beq
\int dx {\g}_{B}(x)^2=1,\
\int dx {\g}_{k_1} (x) {\g}^*_{k_2}(x)=2\pi \delta(k_1-k_2),\ 
\int dx {\g}_{S_1}(x){\g}_{S_2}(x)=\delta_{S_1S_2}. \label{comp}
\eeq
The sign of $\g_B(x)$ is fixed using
\beq
\g_B(x)=-\frac{f\p(x)}{\sqrt{Q_0}}. \label{gb}
\eeq

\subsection{Moving Classical Kink}
The moving kink
\beq
\phi(x,t)=f\gx\hsp \pi(x,t)=\partial_t\phi(x,t)=-v\gamma f\p\gx
\eeq
satisfies the equations of motion
\bea
\ddot\phi(x,t)-\partial_x^2\phi(x,t)+\frac{V^{(1)}(\sl\phi(x,t))}{\sl}&=&(v^2\gamma^2-\gamma^2)f\pp\gx+\frac{V^{(1)}(\sl f\gx)}{\sl}\nonumber\\
&=&-f\pp\gx+\frac{V^{(1)}(\sl f\gx}{\sl}=0.\nonumber
\eea

Its energy is
\bea
E&=&\int dx \left[ 
\frac{\pi^2(x)}{2}+\frac{\left(\partial_x \phi(x)\right)^2}{2}+\frac{V(\sqrt{\lambda} \phi(x))}{\lambda}
\right]\label{en}\\
&=&\int dx \left[ 
\frac{v^2\gamma^2 f^{\prime 2}\gx}{2}+\frac{\gamma^2 f^{\prime 2}\gx}{2}+\frac{V(\sqrt{\lambda} f\gx)}{\lambda}
\right]\nonumber\\
&=&\int \frac{dy}{\gamma} \left[ 
\frac{1+v^2}{1-v^2}\frac{f^{\prime 2}(y)}{2}+\frac{V(\sqrt{\lambda} f(y)}{\lambda}
\right]
=
\int \frac{dy}{\gamma}\frac{f^{\prime 2}(y)}{1-v^2}=Q_0\gamma.
\nonumber
\eea


\section{Quantum Kink at Rest} \label{quantsez}

The normal ordering in Eq.~(\ref{heq}) is defined at the mass of the field $\phi(x)$ near one of the minima of the potential
\beq
m^2=V^{(2)}(\sqrt{\lambda} f(\pm \infty))
.
\eeq
The values obtained at the minima on the two sides of the kink must be equal, or else the kink will begin to accelerate \cite{wstabile}.

\subsection{Ground State Kink}
The defining frame is a choice of identification of the vectors in the Hilbert space, which we denote with kets, with the states in the quantum theory.  In the defining frame, let $|K\rangle$ be vector corresponding to the ground state kink at rest, in other words it is defined to be the Hamiltonian eigenstate with the lowest energy in the kink sector, which consists of states with a single kink and a finite number of mesons.

We may rewrite this state using the unitary displacement operator $\df$
\beq
|K\rangle=\df\vac\hsp 
\df=\exp{-i\int dx f(x) \pi(x)}
\eeq
where $\vac$ is defined to be $\df^\dag|K\rangle$.  In the defining frame, $\vac$ represents a state in the vacuum sector, which consists of states with finite numbers of mesons and no kinks.

In the defining frame, the energy of a state is the corresponding eigenvalue of the Hamiltonian $H$, while time evolution is generated by the action of $H$.  The state $|K\rangle$ is defined to be a Hamiltonian eigenstate
\beq
H|K\rangle=H\df\vac=Q\df\vac=Q|K\rangle. \label{hvac}
\eeq

We define the kink frame to be a different identification of the vectors in the Hilbert space with quantum states.  In particular, a state called $|\Psi\rangle$ in the defining frame is called $\df^\dag|\Psi\rangle$ in the kink frame.  For example, the kink ground state is denoted by $|K\rangle$ in the defining frame but in the kink frame we call it $\vac$.  In summary, the vector $\vac$ represents a vacuum sector state in the defining frame but it represents a kink sector state in the kink frame. 

The transition between the frames is a passive transformation, and so it transforms not only the coordinates on the Hilbert space, but also the operators acting on those coordinates.  For example, it transforms the Hamiltonian into the kink Hamiltonian $H\p$
\beq
H\p=\df^\dag H\df. \label{hpdef}
\eeq
The ket $\vac$ is a kink Hamiltonian eigenstate
\beq
H\p\vac=Q\vac.
\eeq
Note that this follows from Eqs.~(\ref{hvac}) and (\ref{hpdef}), it is a purely algebraic identity independent of the frame which is used to identify $\vac$ with a physical state.

More generally, for any operator $\co$ acting in the defining frame, we may define an operator $\co\p$ acting in the kink frame
\beq
\co\df|\Psi\rangle=\df\co\p|\Psi\rangle\hsp \co\p=\df^\dag\co\df.
\eeq
For example, the ground state kink is translation invariant, which in the defining frame means
\beq
P\df\vac=0\hsp P=\int dx \phi(x)\partial_x \pi(x)
\eeq
and, upon multiplying by $\df^\dag$ becomes the kink frame statement
\beq
P\p\vac=0\hsp P\p=\df^\dag P\df.
\eeq
To evaluate this further, let us decompose the ground state kink rest mass $Q$ in orders of $\lambda$
\beq
Q=\sum_{i=0}^\infty Q_i
\eeq
and, following \cite{cahill76}, decompose the Schrodinger picture fields in terms of the normal modes $\g_i(x)$ that solve the Sturm-Liouville equation (\ref{sl}) with frequency $\omega_i$
\bea
\phi(x) &=&\phi_0 \mathfrak{g}_B(x)+\ppin{k} \left(B_k^{\ddag}+\frac{B_{-k}}{2 \omega_k}\right) \mathfrak{g}_k(x) \label{dec}\\
\pi(x) &=&\pi_0 \mathfrak{g}_B(x)+i \ppin{k}\left(\omega_k B_k^{\ddag}-\frac{B_{-k}}{2}\right) \mathfrak{g}_k(x) \nonumber
\eea
where we adopt the conventions
\beq
B_k^\ddag=\frac{B_k^{\dagger}}{2 \omega_k}\hsp
B_{-S}=B_S.
\eeq
The symbol $\dint$ represents the sum of an integral over continuum modes $k$ with a sum $\sum_S$ over shape modes $S$.

Then, using Eq.~(\ref{gb}), one may calculate
\beq
P\p=\sqrt{Q_0}\pi_0+P
\eeq
where the first term translates the kink center of mass and the second translates the mesons.

There is a perturbative expansion in powers of $\sl$
\beq
\vac=\sum_{i=0}^\infty\vac_i\hsp H\p=\sum_{i=0}^\infty H\p_i 
\eeq
such that
\beq
H\p_0=Q_0\hsp H\p_1=0\hsp H\p_2\vac_0=Q_1\vac_0\hsp \pi_0\vac_0=B_k\vac_0=0.
\eeq
Each successive term in $\vac$ and $H\p$ contains an additional power of $\sl$ while each term in $Q$ contains a power of $\lambda$.   In particular, $Q_0$ is of order $O(1/\lambda)$ and so
\beq
0=P\p\vac=\sqrt{Q_0}\pi_0\vac+P\vac
\eeq
implies
\beq
\pi_0\vac_1=-\frac{1}{\sqrt{Q_0}}P\vac_0.
\eeq
The left hand side is the momentum of the center of mass of the kink while the right hand side is minus the meson momentum.  Of course, these two contributions to the momentum must cancel because we are working in the center of mass frame as $P\p\vac=0$. 

\subsection{Excited Kink}

A kink may be excited by adding a meson or shape mode, which we will refer to formally using the index $k$.  In the kink frame this excited state corresponds to the ket vector $|k\rangle$ and in the defining frame to the vector $\df|k\rangle$.  We demand that it is translation invariant
\beq
P\df|k\rangle=0\hsp P\p|k\rangle=0.
\eeq
It is also required to be a Hamiltonian eigenstate
\beq
H\df|k\rangle=(Q+\ok{}+O(\lambda))\df|k\rangle\hsp H\p|k\rangle=(Q+\ok{}+O(\lambda))|k\rangle.
\eeq
At leading order, in the kink frame, the corresponding ket vector is
\beq
|k\rangle_0=\Bd{}\vac_0\hsp
|k\rangle=\sum_{i=0}^\infty|k\rangle_i
\eeq
and so in the defining frame at leading order it is $\df|k\rangle_0$.  The translation invariance condition fixes the leading correction, up to a term in the kernel of $\pi_0$, to be
\bea
\pi_0|k\rangle_1&=&-\frac{1}{\sqrt{Q_0}}P|k\rangle_0=-\frac{1}{\sqrt{Q_0}}P\Bd{}\vac_0=-\frac{1}{\sqrt{Q_0}}\left([P,\Bd{}]+\Bd{}P\right)\vac_0\\
&=&-\frac{1}{\sqrt{Q_0}}[P,\Bd{}]\vac_0+\Bd{}\pi_0\vac_1.
\eea
In other words, up to corrections of order $O(\sl)$ in the kernel of $\pi_0$ and arbitrary corrections of $O(\lambda)$
\beq
|k\rangle=\left(\Bd{}-\frac{[P,\Bd{}]}{\sqrt{Q_0}}\right)\vac.
\eeq
In particular, this approximation to $|k\rangle$ is sufficient to ensure translation invariance $P\p|k\rangle=0$ at leading order. 

To evaluate the commutator term, let us expand
\beq
\phi(x)=\phi_0\g_B(x)+\ppin{k}\phi_k \g_k(x)\hsp \pi(x)=\pi_0\g_B(x)+\ppin{k}\pi_k \g_k(x).
\eeq
In the case of continuum and shape modes
\beq
\phi_k=\Bd{}+\frac{B_{-k}}{2\ok{}}
\hsp
\pi_k=i\ok{}\Bd{}-i\frac{B_{-k}}{2}.
\eeq
Therefore
\beq
[\phi(x),\Bd{}]=\frac{\g_{-k}(x)}{2\ok{}}\hsp
[\pi(x),\Bd{}]=-i\frac{\g_{-k}(x)}{2}.
\eeq
Assembling the pieces
\bea
[P,\Bd{}]&=&\int dx\left( [\phi(x),\Bd{}]\partial_x\pi(x)- \partial_x\phi(x) [\pi(x),\Bd{}]\right)\\
&=&\frac{1}{2}\Delta_{-k B}\left(\frac{\pi_{0}}{\ok{}}+i\phi_{0}\right)
+\frac{1}{2}\ppin{k\p}\Delta_{-k k\p}\left(\frac{\pi_{k\p}}{\ok{}}+i\phi_{k\p}
\right)\nonumber\\
&=&\frac{1}{2}\Delta_{-k B}\left(\frac{\pi_{0}}{\ok{}}+i\phi_{0}\right)
+\frac{i}{2}\ppin{k\p}\Delta_{-k k\p}\left(\frac{2\okp{}\Bdp{}-B_{-k\p}}{2\ok{}}+\Bdp{}+\frac{B_{-k\p}}{2\okp{}}
\right)
\nonumber\\
&=&\frac{\Delta_{-k B}}{2\ok{}}\left(\pi_{0}+i\ok{}\phi_{0}\right)
+\frac{i}{4\ok{}}\ppin{k\p}\Delta_{-k k\p}\left({2(\okp{}+\ok{})\Bdp{}+\left(\frac{\ok{}}{\okp{}}-1\right)B_{-k\p}}
\right).
\nonumber
\eea

As a result of the time independence of $f(x)$, one has
\beq
 \partial_t \df =\df \partial_t.
\eeq
Therefore, time evolution of any kink frame vector $|\Psi\rangle$ can be easily derived from the defining frame time evolution equation
\beq
\partial_t\df|\Psi\rangle=-iH\df|\Psi\rangle.
\eeq
One finds
\beq
\partial_t|\Psi\rangle=\partial_t\df^\dag\df|\Psi\rangle=\df^\dag \partial_t \df|\Psi\rangle=-i\df^\dag H \df|\Psi\rangle=-iH\p|\Psi\rangle.
\eeq
So we learn that the kink Hamiltonian $H\p$ generates time evolution in the kink frame.

\section{Moving Quantum Kink} \label{movesez}

\subsection{Boost Operator}
In the defining frame, we may boost any state using the boost operator
\beq
\Lambda=-tP[\pi(x),\phi(x)]+\int dx x \ch(\pi(x),\phi(x)).
\eeq
Using
\bea
[P,\phi(x)]&=&-\int dy [\pi(y),\phi(x)]\partial_y\phi(y)=i\partial_x\phi(x)\\
\left[P,\pi(x)\right]&=&\int dy [\phi(y),\pi(x)]\partial_y\pi(y)=i\partial_x\pi(x)\nonumber
\eea
at time $t=0$ one finds
\beq
[P,\Lambda]=i\int dx x\partial_x\ch(\pi(x),\phi(x))=-iH.
\eeq
Also the Poincar\'e algebra implies
\beq
[H,\Lambda]=-iP\hsp [H,P]=0.
\eeq

In the kink frame this becomes
\beq
\Lambda\p=\df^\dag\Lambda \df=\int dx x \ch(\pi(x),\phi(x)+f(x))=\int dx x\ch\p(\pi(x),\phi(x))
\eeq
where $\ch\p$ is the density of the kink Hamiltonian.

\subsection{Moving Ground State Kink}

Consider an eigenstate of the both the Hamiltonian $H$  and also the momentum $P$.  As they mix under boosts
\beq
e^{-i\alpha\Lambda}He^{i\alpha\Lambda}=H\cosh{\alpha}+P\sinh{\alpha}\hsp
e^{-i\alpha\Lambda}Pe^{i\alpha\Lambda}=P\cosh{\alpha}+H\sinh{\alpha}
\eeq
a boost of this state will also be an eigenstate of both $H$ and $P$, but the eigenvalues will change.

\subsubsection{The Defining Frame}

Define the state 
\beq
|K\rangle^v=e^{i\alpha\Lambda}|K\rangle
\eeq
to be the kink ground state boosted to a velocity $v$ or equivalently a rapidity $\alpha=$arctanh$(v)$.  Now the eigenvalues are
\beq
H|K\rangle^v=Q\gamma |K\rangle^v\hsp P|K\rangle^v=Qv\gamma |K\rangle^v.
\eeq
As $|K\rangle^v$ is written in the defining frame, the time evolution is just
\beq
\partial_t|K\rangle^v=-iH|K\rangle^v.
\eeq

\subsubsection{An Inertial Frame}

The inertial frame is defined by the passive transformation $\df$
\beq
|K\rangle^v=\df \vac^v.
\eeq
In the inertial frame, the moving kink state is represented by an eigenvector of the kink Hamiltonian and the kink momentum
\bea
H\p\vac^v&=&H\p\df^\dag|K\rangle^v=\df^\dag H|K\rangle^v=Q\gamma\df^\dag|K\rangle^v=Q\gamma\vac^v\label{hp}\\
P\p\vac^v&=&P\p\df^\dag|K\rangle^v=\df^\dag P|K\rangle^v=Qv\gamma\df^\dag|K\rangle^v=Qv\gamma\vac^v.\nonumber
\eea

Note that $\vac^v$ is not annihilated by $P\p$, indeed it has an eigenvalue proportional to $Q$ which is of order $O(1/\lambda)$, potentially mixing various orders in perturbation theory when this condition is imposed.  

To see that this is reasonable, let us try to understand the state $\vac^v$ more explicitly.  It is
\beq
\vac^v=\df^\dag|K\rangle^v=\df^\dag e^{i\alpha\Lambda}|K\rangle=e^{i\alpha\Lambda\p}\df^\dag|K\rangle=e^{i\alpha\Lambda\p}\vac.
\eeq
We may expand $\Lambda\p$ in powers of the coupling
\bea
\Lambda\p&=&\sum_{i=0}^\infty \Lambda\p_i\hsp
\Lambda\p_0=0\label{poteri}\\
\Lambda\p_1&=&\int dx x \left[\partial_x\phi(x)\partial_x f(x) +\frac{\V1}{\sl} \phi(x) \right]\nonumber\\
&=&\int dx \phi(x)\left( x \left[-\partial^2_x f(x) +\frac{\V1}{\sl}  \right]-\partial_x f(x)\right)\nonumber=\sqrt{Q_0}\phi_0\nonumber\\
\Lambda\p_2&=&\frac{1}{2}\int dx x \left[\pi^2(x)+\left(\partial_x\phi(x)\right)^2+\V2 \phi^2(x)\right].\nonumber
\eea

Recall that the semiclassical limit requires $\alpha\ll 1$.  Let us therefore take the limit $\alpha\rightarrow 0$ but we do not impose that $\alpha/\sqrt{\lambda}\rightarrow 0$.  In this limit
\bea
e^{i\alpha\Lambda\p}&=&e^{i\alpha\Lambda\p_1}+i\int_0^\alpha d\beta e^{i(\alpha-\beta)\Lambda\p_1}\Lambda\p_2 e^{i\beta\Lambda\p_1}\\
&=&e^{i\alpha\sqrt{Q_0}\phi_0}+i\int_0^\alpha d\beta e^{i(\alpha-\beta)\sqrt{Q_0}\phi_0}\Lambda\p_2 e^{i\beta\sqrt{Q_0}\phi_0}.\nonumber
\eea
Using
\beq
[\pi_0,e^{i\theta\phi_0}]=\theta e^{i\theta\phi_0}\hsp
[\pi(x),e^{i\theta\phi_0}]=\theta\g_B(x) e^{i\theta\phi_0}
\eeq
one finds
\beq
e^{i(\alpha-\beta)\sqrt{Q_0}\phi_0}\pi^2(x) e^{i\beta\sqrt{Q_0}\phi_0}=e^{i\alpha\sqrt{Q_0}\phi_0}\left(\pi(x)+\beta\sqrt{Q_0}\g_B(x)\right)^2.
\eeq
The $\beta^2$ term leads to an $x$ integral proportional to $x\g_B^2(x)$ which is odd, and so it vanishes.  The linear term, when integrated over $\beta$, is proportional to $\alpha^2$, which we drop as we are working at linear order in $\alpha$.  In all, the Lorentz transformation is
\beq
e^{i\alpha\Lambda\p}=e^{-i\alpha\sqrt{Q_0}\phi_0}\left(1+i\alpha\Lambda\p_2\right). \label{lt}
\eeq

Our inertial frame ground state kink is then
\beq
\vac^v=e^{i\alpha\sqrt{Q_0}\phi_0}\left(1+i\alpha\Lambda\p_2\right)\vac.
\eeq
Now 
\bea
P\p\vac^v&=&\left(\sqrt{Q_0}\pi_0+P\right)e^{i\alpha\sqrt{Q_0}\phi_0}\left(1+i\alpha\Lambda\p_2\right)\vac\\
&=&(Q_0\alpha+P)\vac^v +\sqrt{Q_0}e^{i\alpha\sqrt{Q_0}\phi_0}\pi_0\left(1+i\alpha\Lambda\p_2\right)\vac.\nonumber
\eea
The $Q_0\alpha$ term is of order $O(1/\lambda)$ and it exactly reproduces the eigenvalue in Eq.~(\ref{hp}).  Therefore this nonvanishing eigenvalue is simply a result of the $e^{i\alpha\sqrt{Q_0}\phi_0}$ coefficient in the inertial frame state.  It can be factored out, and then the usual perturbation theory derivation of the states follows with $P\p$ annihilating the rest of the state.  This is done automatically in the comoving frame.

\section{The Comoving Frame} \label{cosez}

\subsection{Definition}

The comoving frame is defined using the passive transformation
\bea
\dft&=&\exp{-i\int dx \left(f\gx\pi(x)-\partial_t f\gx \phi(x)\right)}\\
&=&\exp{-i\int dx \left( f\gx\pi(x)+\gamma v f\p\gx \phi(x)\right)}\nonumber
\eea
which shifts $\phi(x)$ and $\pi(x)$ by the classical moving kink solution.  It is sometimes convenient to regard $t$ as a parameter, so that one comoving frame is introduced for each value of $t$.  One can use the comoving frame in which $t$ is equal to the time, but that choice is not necessary.

Consider a state that is represented by the ket $|\Psi\rangle$ in the defining frame.  In the comoving frame, this state is represented by the ket $|\Psi\rangle^{(t)}$
\beq
|\Psi\rangle^{(t)}=\dfd|\Psi\rangle.
\eeq

\subsection{The Ground State Kink}

In particular, in the comoving frame, we write the moving ground state kink as
\beq
\vac^{(t)v}=\dfd|K\rangle^v=\dfd\df\vac^v=\dfd\df e^{i\alpha\Lambda\p}\vac.
\eeq
Using the Baker-Campbell-Hausdorff formula we may evaluate
\bea
\dfd\df&=&\exp{i\int dx \left((f\gx-f(x))\pi(x)-\dot f\gx \phi(x)\right)}\nonumber\\
&&\times\exp{-i\int dx \frac{\dot f\gx f(x)}{2}}. \label{dfdf}
\eea
Note that the last factor is an overall constant phase.  This may be simplified using
\beq
-i\int dx \dot f\gx = iv\gamma\int dx f\p\gx=-iv\gamma \sqrt{Q_0}\phi_0=-i\ \sinh{\alpha}\sqrt{Q_0}\phi_0.
\eeq
Thus the $\phi(x)$ term in (\ref{dfdf}) cancels the $e^{i\alpha\Lambda\p_1}=e^{-i\alpha\sqrt{Q_0}\phi_0}$ term in $e^{i\alpha\Lambda\p}$, reported in Eq.~(\ref{lt}), up to corrections of order $\alpha^3\sqrt{Q_0}$, which we drop as higher powers of $\alpha$ arise at higher orders of perturbation theory.

Assembling these results, we find
\beq
\dfd\df e^{i\alpha\Lambda\p}=\exp{i\int dx \left(\left[f\gx-f(x)\right]\pi(x)-\frac{\dot f\gx f(x)}{2}\right)}
\left(1+i\alpha\Lambda\p_2\right) \label{otrans}
\eeq
and so in the comoving frame, the kink ground state corresponds to the ket
\beq
\vac^{(t)v}=\exp{i\int dx \left(\left[f\gx-f(x)\right]\pi(x)-\frac{\dot f\gx f(x)}{2}\right)}
\left(1+i\alpha\Lambda\p_2\right)\vac.
\eeq
We see that all that remains at order $O(1/\lambda)$ is a constant phase, which will be inconsequential and we will drop it from now on.  

At order $O(1/\sl)$ one finds a $\pi(x)$ tadpole term that Lorentz contracts and moves the boosted kink.  At time $t=0$ it does not move the boosted kink.  We are free to choose any time to be $t=0$, as this choice simply corresponds to the choice of kink position in $\df$ which does not affect the state.  The contraction appears only at order $O(v^2/\sl)$.

\subsection{Evolution}

Evolution in the defining frame is generated by $H$.  We will now use this fact to learn how to evolve kets in the comoving frame. 

Evolution in the comoving frame is described by
\beq
\partial_t\vac^{(t)v}=\partial_t\dfd|K\rangle^v=[\partial_t,\dfd]|K\rangle^v+\dfd\partial_t|K\rangle^v. \label{com}
\eeq

The second term is easily evaluated, as it corresponds to evolution in the defining frame
\beq
\dfd\partial_t|K\rangle^v=-i\dfd H|K\rangle^v=-i \dfd H\dft\vac^{(t)v}.
\eeq
To evaluate the first term
\beq
[\partial_t,e^{i\int dx \left(f\pi-\dot f\phi\right)}]\dft\vac^{(t)v}
\eeq
on the right hand side of (\ref{com}), use 
\beq
[\partial_t,f\pi-\dot f\phi]=\dot f\pi-\ddot f\phi
\eeq
to find
\bea
[\partial_t,e^{i\int dx \left(f\pi-\dot f\phi\right)}]&=&\sum_{n=0}^\infty \frac{i^n}{n!}\left[
\partial_t,
\left(\int dx f\pi-\dot f\phi\right)^n
\right]\\
&&\hspace{-2cm}=\sum_{n=1}^\infty \frac{i^n}{n!}\sum_{m=1}^n 
\left(\int dx f\pi-\dot f\phi\right)^{m-1}
\left(\int dx \dot f\pi-\ddot f\phi\right)
\left(\int dx f\pi-\dot f\phi\right)^{n-m}\nonumber\\
&&\hspace{-2cm}=\left(\int dx \dot f\pi-\ddot f\phi\right)
\sum_{n=1}^\infty \frac{i^n}{n!}\sum_{m=1}^n 
\left(\int dx f\pi-\dot f\phi\right)^{n-1}
\nonumber\\
&&\hspace{-2cm}\ \ +\sum_{n=1}^\infty \frac{i^n}{n!}\sum_{m=1}^n 
\left[\left(\int dx f\pi-\dot f\phi\right)^{m-1}
,\left(\int dx \dot f\pi-\ddot f\phi\right)\right]
\left(\int dx f\pi-\dot f\phi\right)^{n-m}\nonumber\\
&&\hspace{-2cm}=\left(\int dx \dot f\pi-\ddot f\phi\right)
\sum_{n=1}^\infty \frac{i^n}{(n-1)!}
\left(\int dx f\pi-\dot f\phi\right)^{n-1}
\nonumber\\
&&\hspace{-2cm}\ \ +\sum_{n=1}^\infty \frac{i^n}{n!}\sum_{m=1}^n 
(m-1)\int dx(if\ddot f-i\dot f^2)
\left(\int dx f\pi-\dot f\phi\right)^{m-2}\left(\int dx f\pi-\dot f\phi\right)^{n-m}
\nonumber\\
&&\hspace{-2cm}=i\int dx (\dot f\pi-\ddot f\phi)e^{i\int dx f\pi-\dot f \phi}-\sum_{n=2}^\infty \frac{i^{n-1}}{(n-2)!}
\left(\int dx f\pi-\dot f\phi\right)^{n-2}
\int dx\frac{f\ddot f-\dot f^2}{2}
\nonumber\\
&&\hspace{-2cm}=-i \int dx\left(\ddot f\phi-\dot f\pi +\frac{f\ddot f-\dot f^2}{2}\right)\dfd\nonumber.
\eea
Combining these terms one finds
\beq
\partial_t\vac^{(t)v}=-i\hp\vac^{(t)v}
\eeq
where the comoving Hamiltonian is defined to be
\bea
\hp&=& \dfd H\dft+\int dx\left(\ddot f\gx\phi(x)-\dot f\gx\pi(x)\right.\\
&&\left.+\frac{f\gx\ddot f\gx-\dot f^2\gx}{2}\right).\nonumber
\eea
This comoving Hamiltonian evolves the state while changing the comoving frame parameter $t$ so that it equals the time.  

Again we decompose it in powers $\hp_i$ of the coupling and evaluate it
\bea
\hp[\phi(x),\pi(x)]&=&H[\phi(x)+f\gx,\pi(x)+\dot f\gx]+\int dx\left(
\ddot f\gx\phi(x)
\right.\nonumber\\
&&\left.-\dot f\gx\pi(x)+\frac{f\gx\ddot f\gx-\dot f^2\gx}{2}\right)\nonumber\\
&=&\sum_{i=0}^\infty \hp_i.
\eea
The first term is
\beq
\hp_0=\int dx\left[\frac{f\gx\ddot f\gx+\left(\partial_x f\gx\right)^2}{2}+V(\sl f\gx)\right].
\eeq
Unlike Eq.~(\ref{en}), the result is not $Q_0\gamma$, as a result of the $f\ddot f-\dot f^2$ term which negates the kinetic energy.  While the commutator term in (\ref{com}) makes $\hp_0$ more complicated, it also exactly cancels the tadpole in $\dfd H\dft$
\beq
\hp_1=0.
\eeq
It does not contain terms at higher orders, so these are given by $\dfd H\dft$.  For example, leaving the normal ordering implicit,
\beq
\hp_2=\frac{1}{2}\int dx\left[ 
\pi^2(x)+(\partial_x\phi(x))^2+V^{(2)}(\sl f\gx)\phi^2(x)\right].
\eeq

\subsection{Operators}

Refs.~\cite{cahill76,me2loop} construct the spectrum of a stationary kink in the kink frame using kink frame operators.  In this subsection, we will see how to map from these operators to the comoving frame of a moving kink, so that the old construction may be imported to the case of a moving kink.

Consider an operator
\beq
\co\p=\df^\dag\co\df
\eeq
acting on a stationary kink state $|\Psi\rangle$\ in the kink frame.  How does it act on the corresponding boosted kink in the comoving frame?  

After the action of the operator, in the kink frame the state is represented by the vector
\beq
\co\p|\Psi\rangle=\df^\dag\co\df|\Psi\rangle
\eeq
which, in the defining frame, corresponds to
\beq
\co\df|\Psi\rangle.
\eeq
Now we may boost it, yielding a new state which in the defining frame is
\beq
e^{i\alpha\Lambda}\co\df|\Psi\rangle. \label{act}
\eeq
In the comoving frame, before acting with the operator the boosted state was
\beq
|\Psi\rangle^{v(t)}=\dfd e^{i\alpha\Lambda}\df|\Psi\rangle=\dfd\df e^{i\alpha\Lambda\p}|\Psi\rangle.
\eeq
Acting with the operator it becomes the state (\ref{act}) which in the comoving frame is represented by
\beq
\dfd e^{i\alpha\Lambda}\co\df|\Psi\rangle=
\dfd\df e^{i\alpha\Lambda\p}\co\p|\Psi\rangle=\co^{\prime v (t)}|\Psi\rangle^{v(t)}
\eeq
where
\beq
\co^{\prime v (t)}=\dfd\df e^{i\alpha\Lambda\p}\co\p e^{-i\alpha\Lambda\p} \df^\dag\dft.
\eeq
Using Eq.~(\ref{otrans}) this yields our master formula
\beq
\co^{\prime v (t)}=e^{i\int dx \left(f\gx-f(x)\right)\pi(x)}
\left(\co\p+i\alpha[\Lambda\p_2,\co\p]\right)e^{-i\int dx \left(f\gx-f(x)\right)\pi(x)}. \label{padrone}
\eeq
For any operator $\co\p$ acting on a stationary kink in the kink frame, we have found the corresponding operator $\co^{\prime v (t)}$ that acts on the moving kink in the comoving frame.  This formula can be used to migrate all results concerning a stationary kink to a moving kink.  For example, we know how to build the spectrum of a stationary kink using the operators $B^\ddag$, $B$, $\phi_0$ and $\pi_0$ in the kink frame, and so this map yields the corresponding construction for a moving kink in the comoving frame.

Recall that, choosing $t=0$, the Lorentz-contracting $\pi(x)$ terms are of order $O(v^2/\sl)$ and so may be ignored at lower orders.  Therefore, at these low orders, for any stationary operator $\co\p$ in the kink frame of a stationary kink, the operator
\beq
\co^{\prime v (t)}=\co\p+i\alpha[\Lambda\p_2,\co\p] \label{app}
\eeq
acts identically on the same kink state, but boosted and in the comoving frame.  Conversely, for any operator $\co$ acting on a moving kink in the comoving frame, the operator 
\beq
\hat\co=\co-i\alpha[\Lambda\p_2,\co] \label{inv}
\eeq
acts identically on the kink in its rest Lorentz frame in the kink frame.  We remind the reader that the two uses of the word ``frame" are distinct.  One refers to a Lorentz frame and the other to the identification between kets and states.

\subsection{Fields}
Let us apply the transformation (\ref{inv}) to the scalar field $\co=\phi(x)$ acting in the comoving frame of a moving kink.  The action of the field $\phi(x)$ in the comoving frame of the moving kink corresponds to that of
\beq
\hat\phi(x)=\phi(x)-[i\alpha\Lambda\p_2,\phi(x)]=\phi(x)-\alpha x\pi(x)
\eeq
in the kink frame of a stationary kink, where we evaluated the commutator using the explicit expression for $\Lambda\p_2$ in Eq.~(\ref{poteri}).  In other words, the operator $\hat\phi(x)$ acting on a state of a stationary kink in the kink frame changes it to the same state, up to a boost, as one would get by acting on the same state but boosted with $\phi(x)$, using the comoving frame.

Similarly, acting with the operator $\pi(x)$ in the comoving frame has the same effect on the state as acting on the stationary kink with
\bea
\hat\pi(x)&=&\pi(x)-[i\alpha\Lambda\p_2,\pi(x)]=\pi(x)+\alpha \left(-\partial_x^2+V^{(2)}(\sl f\gx) 
\right)(x\phi(x))\nonumber\\
&=&\pi(x)+\alpha \ppin{k}\phi_k\left(-\partial_x^2+V^{(2)}(\sl f\gx) \right)(x\g_k(x))
\eea
in the kink frame.  Again dropping terms of order $O(v^2/\sl)$, the Lorentz contraction on the potential term can be dropped and at time $t=0$ we find the Sturm-Liouville operator acting on the normal modes, and so
\beq
\hat\pi(x)=\pi(x)+\alpha
\left(-\partial_x\phi(x)+x\ppin{k}\ok{}^2\g_k(x)\phi_k\right).
\eeq

These equations took a long time to derive, but in the Lagrangian formulation they resemble the usual Lorentz transformation where $\pi(x,t)=\dot\phi(x,t)$ and the Heisenberg equations of motion are satisfied for the comoving kink Hamiltonian.

Recall that in the kink frame of the stationary kink, the operators $B^\ddag$ and $B$ are creation and annihilation operators for mesons.  The operators $B^{\ddag\prime}$ and $B\p$ that play the same role in the comoving frame of the moving kink\footnote{These operators depend on the rapidity and the time, but we will leave this dependence implicit to avoid clutter.} are more complicated to write explicitly, as one needs to add a commutator with respect to $\Lambda\p_2$ following the prescription in (\ref{app}).   

Nonetheless, we may decompose our operators $\phi(x)$ and $\pi(x)$ into $B^{\ddag\p}$ and $B\p$ to learn how the fields are related to the operators that create and destroy mesons in the comoving frame.  To do this, one recalls that $\phi(x)$ and $\pi(x)$ in the comoving frame act identically to $\hat\phi(x)$ and $\hat\pi(x)$ in the kink frame, which we may write in terms of $\phi(x)$ and $\pi(x)$ using the results above, and then decompose into $B^\ddag$ and $B$ as usual.  Then, by definition, the action in the comoving frame is given by replacing these with $B^{\ddag\p}$ and $B\p$.  

Let us run through that argument more slowly and concretely.  Consider the operator $\phi(x)$.  The operators themselves, and the relations between them, are independent of the frame.  However it will be convenient to think of this operator acting on a moving kink state in the comoving frame.  Now, setting $\co\p=\phi(x)$ in (\ref{app}) we have defined another operator $\co^{\prime v(t)}$.  Let us give it another name
\beq
\phi\p(x)=\co^{\prime v(t)}=\phi(x)+i\alpha[\Lambda\p_2,\phi(x)].
\eeq
This operator, acting on the moving kink in the comoving frame, changes the state identically to the operator $\phi(x)$ acting on the kink in its rest frame, using the kink frame.  Similarly, we can also use (\ref{app}) to construct $\pi\p(x)$ which acts in the comoving frame as $\pi(x)$ acts in the kink frame.

We now have three operators: $\phi(x)$, $\phi\p(x)$ and $\pi\p(x)$ which have simple interpretations in the comoving frame.  How are they related?  At leading order if $\alpha$,
\beq
\phi(x)=\phi\p(x)-[i\alpha\Lambda\p_2,\phi\p(x)]=\phi\p(x)-\alpha x\pi\p(x). \label{phipp}
\eeq
Running the same argument with $\pi(x)$ instead of $\phi(x)$ one obtains
\beq
\pi(x)=\pi\p(x)+
\alpha\left(-\partial_x\phi\p(x)+x\ppin{k}\ok{}^2\g_k(x)\phi\p_k\right). \label{pipp}
\eeq

Our goal is not to write $\phi(x)$ and $\pi(x)$ in terms of $\phi\p(x)$ and $\pi\p(x)$, but rather to write them in terms of the operators $B\p$ and $B^{\ddag\prime}$ that destroy and create mesons and shape modes in the comoving frame.  To do this, let us run the decomposition (\ref{dec}) through or master formula (\ref{app}) to produce the comoving frame version
\bea
\phi\p(x) &=&\phi\p_0 \mathfrak{g}_B(x)+\ppin{k} \left(B_k^{\ddag\prime}+\frac{B\p_{-k}}{2 \omega_k}\right) \mathfrak{g}_k(x)\\
\pi\p(x) &=&\pi\p_0 \mathfrak{g}_B(x)+i \ppin{k}\left(\omega_k B_k^{\ddag\prime}-\frac{B\p_{-k}}{2}\right) \mathfrak{g}_k(x). \nonumber
\eea
Plugging this into Eqs.~(\ref{phipp}) and (\ref{pipp}) we find
\bea
\phi(x)&=&
\label{deco}
(\phi\p_0-\alpha x\pi\p_0)\g_B(x)+\ppin{k}\g_k(x)\left[ 
\left(1-i\alpha x\ok{} \right)B_k^{\ddag\prime}{}
+\left(1+i\alpha x\ok{}  \right) \frac{B\p_{-k}}{2\ok{}}
\right]\nonumber\\
\pi(x)
&=&\pi\p_0\g_B(x)-\alpha\phi\p_0\partial_x\g_B(x)+\ppin{k}\left[-\alpha\partial_x\g_k(x)\left(B^{\ddag\prime}_k+ \frac{B\p_{-k}}{2\ok{}}
\right)\right.\nonumber\\
&&\left.
+\ok{}\g_k(x)\left[ 
\left(i+\alpha x\ok{}\right)B^{\ddag\prime}_k{}
+\left(-i+\alpha x \ok{}\right) \frac{B\p_{-k}}{2\ok{}}
\right]\right].
\eea
What does this formula mean?  The fields $\phi(x)$ and $\pi(x) $ are always the same operators, independent of Lorentz frames and frames identifying states with kets.  However, the operators that interpolate between eigenstates of the free Hamiltonian, creating and destroying mesons and translating the center of mass, do depend on the frame.  We have now presented the decomposition of $\phi(x)$ and $\pi(x)$ in terms of those operators in the comoving frame.  It can be applied to perturbative calculations in the comoving frame, as the comoving Hamiltonian is a functional of $\phi(x)$ and $\pi(x)$.

Notice that the expansion in the rapidity $\alpha$ is in fact an expansion in $\alpha x\ok{}$.  Let $\ok{}$ be of order $O(m)$, corresponding to mesons that are relativistic but not ultrarelativistic.  Physical effects such as loop corrections to the kink mass are indeed dominated by this range of $k$.  Then this approximation is only valid at $|x|\ll 1/(\alpha m)$.  The width of the classical kink is $1/m$, and so the approximation is valid over a range of $1/\alpha$ classical kink widths.  Recall that we are interested in nonrelativistic kinks, as perturbation theory is only expected to apply in this case, and so $\alpha\rightarrow 0$.  Therefore this maximum distance can be larger than many other characteristic scales in the problem. 

\subsection{The Main Lesson from this Example}

We are interested in generalizing these results to other time-dependent solutions.  The key to this generalization will be the following observation.  While our derivation was rather tortuous, the decomposition (\ref{deco}) can be simply derived as follows.  One starts with the classical field theory solution for a small perturbation of a stationary kink in the inertial frame, which, as the kink is stationary, coincides with the kink frame $\phi(x,t)\rightarrow \phi(x,t)-f(x)$
\beq
\phi(x,t)=\phi_0 \g_B(x) +\pi_0 t\g_B(x)+ \ppin{k}\g_k(x)\left(\Bd{}e^{i\ok{} t}+\frac{B_{-k}}{2\ok{}}e^{-i\ok{} t}\right).
\eeq
Here $\phi_0,\ \pi_0,\ B^\ddag$\ and $B$ are $c$-number parameters defining the solution.   In quantum field theory this could be interpreted as an interaction picture field. 

Then Lorentz boost, yielding a new solution to the classical equations of motion (\ref{eom}), where we separate the perturbations from the boosted kink using the transformation $\phi(x,t)\rightarrow \phi(x,t)-f(\gamma(x-vt))$
\bea
\phi(x,t)&=&\phi_0 \g_B\gx +\pi_0 (\gamma(t-vx))\g_B\gx\label{clas}\\
&&+ \ppin{k}\g_k\gx\left(\Bd{}e^{i\ok{} (\gamma(t-vx))}+\frac{B_{-k}}{2\ok{}}e^{-i\ok{} (\gamma(t-vx))}\right).\nonumber
\eea
The decompositions of the quantum fields $\phi(x)$ and $\pi(x)$ follow from the rule
\beq
\phi(x)=\phi(x,0)\hsp \pi(x)=\dot\phi(x,0)
\eeq
applied to the classical solution (\ref{clas}).  In particular, one finds Eq.~(\ref{deco}) expanding to linear order in $v$.

\subsection{The Ground State Revisited}

In the kink frame of a ground state kink, the leading approximation $\vac_0$ to the ground state $\vac$ is annihilated by $\pi_0$ and the operators $B_k$.  In the comoving frame of a moving kink, these operators correspond to 
\beq
\pi\p_0=\int dx \pi\p(x)\g_B(x)=\pi_0+\alpha\int dx \partial_x\phi(x)\g_B(x)=\pi_0+\alpha\ppin{k}\Delta_{Bk}\phi_k
\eeq
and
\bea
B\p_{-k}&=&\int dx\g^*_{-k}(x)\left(\ok{}\phi\p(x)+i\pi\p(x)\right)\\
&=&B_{-k}+\alpha\int dx \g^*_{-k}(x) \left(x (\ok{}\pi(x)-i\ppin{k\p}\okp{}^2\g_{k\p}(x)\phi_{k\p})+i\partial_x\phi(x)\right)\nonumber\\
&=&B_{-k}+\alpha\int dx \g^*_{-k}(x) \ppin{k\p} \left[\pi_{k\p}x\ok{}-i\phi_{k\p}(x\okp{}^2-\partial_x)\right] \g_{k\p}(x)\nonumber\\
&=&B_{-k}+i\alpha\int dx \g^*_{-k}(x) \ppin{k\p} \left[\left(\okp{}\Bdp{}-\frac{B_{-k\p}}{2}\right)x\ok{}-\left(\Bdp{}+\frac{B_{-k\p}}{2\okp{}}
\right)(x\okp{}^2-\partial_x)\right] \g_{k\p}(x)\nonumber\\
&=&B_{-k}+i\alpha\int dx \g^*_{-k}(x) \ppin{k\p} \left[\Bdp{}\left(x\okp{}(\okp{}+\ok{})-\partial_x \right)+\frac{B_{-k\p}}{2}\left(x(\okp{}-\ok{})-\frac{\partial_x}{\okp{}}
\right)
\right] \g_{k\p}(x)\nonumber\\
&=&B_{-k}+i\alpha \ppin{k\p} \left[\Bdp{}\left(\hat\Delta_{k k\p}\okp{}(\okp{}+\ok{})-\Delta_{k k\p} \right)+\frac{B_{-k\p}}{2}\left(\hat\Delta_{k k\p}(\okp{}-\ok{})-\frac{\Delta_{k k\p}}{\okp{}}
\right)
\right]\nonumber
\eea
where
\beq
\hat\Delta_{kk\p}=\int dx x\g_k(x)\g_{k\p}(x).
\eeq
One can check that $\pi\p_0$ and $B\p_{-k}$ indeed annihilate the leading order comoving frame ground state $\left(1+i\alpha\Lambda\p_2\right)\vac_0$.

\subsection{General States}

There are two kinds of questions that one may ask.  The first is the initial value problem, in which one starts with an initial state at time $t=0$ and asks what state arises at time $t$.  This can, in principle, be solved in the comoving frame as time evolution is generated by $\hp$ which has been constructed above.

One may also ask the spectrum.  There are several distinct questions here.  The eigenstates of $\hp$ in the comoving frame are time-independent in the comoving frame, in the sense that the ket that describes the state in the comoving frame is the same at all times.  However, although the ket itself is time-independent, the state to which it corresponds depends on $t$.  Correspondingly, the defining frame kets are time-dependent.  The operator $\hp$, except for scalar terms, begins at order $O(\lambda^0)$ and so in principle this problem can be treated perturbatively.

On the other hand, one may ask about states that are truly time-independent, such as the breather Hamiltonian eigenstates in the quantum Sine-Gordon model.  These are states that are eigenstates of the Hamiltonian $H$ in the defining frame, and so by $\dfd H\dft$ in the comoving frame.  The operator $\dfd H\dft$ has tadpole terms of order $O(1/\sl)$
\beq
\dfd H\dft=\int dx \left( \dot f\gx \pi(x) -\ddot f\gx \phi(x)\right)+O(\lambda^0).
\eeq
In the case of the moving kink, the $\ddot f$ term is of order $O(v^2/\sl)$ and so, if $v^2\ll O(\sl)$, it can be treated perturbatively.  

Dropping it, we find
\beq
\dfd H\dft=v \gamma \sqrt{Q_0 } \pi_0 +O(\lambda^0).
\eeq
One may recognize this tadpole as the part of the momentum operator $P\p$ corresponding to the kink center of mass, which was mixed with the Hamiltonian by the boost.  In particular, {\it{if the state was translation-invariant before the boost}}, so that it was annihilated by $P\p$ up to terms of order unity, then after the boost this term will be of order $O(1)$ because $\sqrt{Q_0}\pi_0$ on such a state is $P$  on the state, and $P$ is of order $O(1)$.  Therefore, for such states, there are no terms in $\dfd H\dft$ of order the inverse coupling and we conclude that the spectrum of translation-invariant states can be found perturbatively.

Physically this is reasonable.  Before the boost, translation-invariance means that one starts in a flat superposition of states with different kink positions $x_0$.  After the boost, evolving in time, the kink moves, and so $x_0$ shifts.  However, if the initial wave function of $x_0$ was constant, so that all kink positions were weighted equally, then this shift leaves the state invariant, and so the boosted state remains time-independent.  Of course, these are the kinds of states that we focused upon and often constructed in previous papers.
  
More generally, in the case of time-dependent solitons that are not simply moving but also have some internal dynamics, this tadpole term always appears, reflecting the fact that the quantum state changes as a result of the classical evolution.  But also in this general case, we expect that this evolution is trivial if the quantum state is a flat superposition over the entire classical trajectory.  This will be manifested by the fact that the tadpole term will annihilate the state, and so one need only solve the eigenvalue equation for the perturbative part of $\dfd H\dft$.  For example, in the case of breathers and oscillons, we expect to be able to obtain the spectra perturbatively if we restrict our attention to states that are flat quantum superpositions over the coherent states corresponding to each classical configuration that appears over a period of their evolution.

\section{A General Time-Dependent Solution} \label{gensez}

Finally we are ready to take the lessons learned from the rigidly moving kink and apply them to a general, time-dependent solution of the classical equations of motion.

\subsection{Perturbations}

Consider a time-dependent classical solution $\phi(x,t)=f(x,t)$ of Eq.~(\ref{eom}).  Now let us consider small perturbations $\g(x,t)$
\beq
\phi(x,t)=f(x,t)+\g(x,t).
\eeq
The classical equations of motion are
\beq
0=\ddot\g(x,t)-\partial_x^2\g(x,t)+\frac{V^{(1)}(\sl(f(x,t)+\g(x,t)))-V^{(1)}(\sl(f(x,t)))}{\sl}.
\eeq
At linear order in $\g$ this becomes
\beq
0=\ddot\g(x,t)-\partial_x^2\g(x,t)+V^{(2)}(\sl(f(x,t)))\g(x,t). \label{geq}
\eeq

We are not always interested in Lorentz-invariant theories.  For example, kink-impurity scattering is phenomenologically rich and tractable with our methods, and yet the impurity necessarily breaks spatial translation symmetry and may even break time translation symmetry.  

However, in the case of a Lorentz-invariant theory, two solutions $\g$ are always present.  These are the spatial and temporal zero modes
\beq
\g_B(x,t)=c_B \partial_x f(x,t)\hsp \g_T(x,t)=c_T \partial_t f(x,t)
\eeq
corresponding to a spatial translation of the solution or a displacement of the solution along its trajectory.  Here $c_B$ and $c_T$ are normalization constants which may, at this point, be chosen freely.

The semiclassical treatment of solitons only is a reasonable approximation for heavy solitons, whose mass is much greater than any momentum scale.  They are necessarily nonrelativistic, but in a Lorentz-invariant theory there will be solutions related by Galilean invariance.  These correspond to the solutions
\beq
\g_V(x,t)=c_V t \g_B(x,t).
\eeq
Formally one may worry that the linear expansion may not be trusted at large $|t|$.  However, for the purpose of quantizing a soliton, $t$ can be chosen freely, as a shift in $t$ corresponds to a shift in the choice of $f(x,t)$ in the moduli space of solutions.

In addition to these three solutions, consider a set of solutions $\g_k(x,t)$ where the index $k$ will in general contain a continuum corresponding to real values of $k$ and also discrete shape modes $S$.  This set of solutions must be linearly independent, but also must span the possible initial perturbations.   

More precisely, consider the pairs
\beq
(\g_T(x,0),\dot\g_T(x,0)),\ (\g_B(x,0),\dot\g_B(x,0)),\ (\g_V(x,0),\dot\g_V(x,0)),\ (\g_k(x,0),\dot\g_k(x,0)).
\eeq
We demand that these pairs are linearly independent and also span the space of pairs of functions $(G(x,0),\dot G(x,0))$. Note that $\g_V(x,0)=0$ but $\dot\g_V(x,0)$ is nonzero and so is not an obstruction to this condition.  In the case of a time-independent solution, $\g_T$ vanishes, but we are not interested in that case as it has been covered in the literature.

In practice, one needs to simply search for solutions to (\ref{geq}) until these span the space of functions, throwing away solutions with linearly dependent initial data.

\subsection{The Comoving Frame}

The displacement operator is defined to be
\beq
\dft=\exp{-i\int dx \left(f(x,t)\pi(x)-\partial_t f(x,t) \phi(x)\right)}.
\eeq
The state $|\Psi\rangle^{(t)}$ in the comoving frame is defined to be the same state as $\dfd|\Psi\rangle$ in the defining frame.

Now the arguments in Sec.~\ref{cosez} may be imported, as the derivations are generally identical.  In particular, time evolution is generated by the comoving Hamiltonian
\bea
\hp&=& \dfd H\dft+\int dx\left(\ddot f(x,t)\phi(x)-\dot f(x,t)\pi(x)
+\frac{f(x,t)\ddot f(x,t)-\dot f^2(x,t)}{2}\right)\nonumber
\eea
while time-independent states are eigenstates of the inertial Hamiltonian $\dfd H\dft$.  In particular
\beq
\hp_1=0\hsp \hp_2=\frac{1}{2}\int dx\left(\pi^2(x)+(\partial_x\phi(x))^2+V^{(2)}(\sl f(x,t))\phi^2(x)\right). \label{hpe}
\eeq

Note that $\g_B,\ \g_T$\ and $\g_V$ are real, while the space of $\g_k$ is invariant under complex conjugation.  

\subsection{The Decomposition}

Following the example of the moving kink, for each value of the parameter $t$ we introduce the decomposition
\bea
\phi(x)&=&\g_B(x,0)\phi_0+\g_T(x,0)\phi_T+\ppin{k}\left(\g_k(x,0)\Bd{}+\g^*_k(x,0)\frac{B_{k}}{2\ok{}}\right)\label{gendec}\\
\pi(x)&=&\dot\g_B(x,0)\phi_0+\dot\g_T(x,0)\phi_T+\g_B(x,0)\pi_0\nonumber
+\ppin{k}\left(\dot\g_k(x,0)\Bd{}+\dot\g^*_k(x,0)\frac{B_{k}}{2\ok{}}\right).\nonumber
\eea
At this point $\ok{}$ is arbitrary, although we define $B^\ddag=B^\dag/(2\ok{})$.  It is included to allow us to simplify expressions in applications.  Note that as a result $B^\ddag$ and $B$ do not necessarily satisfy an oscillator algebra, however their algebra may derived by inverting (\ref{gendec}) and using the canonical commutation relations for $\phi(x)$ and $\pi(x)$.  We will argue in a moment that, although the operators $B^\ddag$ and $B$ themselves implicitly depend on time, the resulting algebra is independent of time.

Using Eq.~(\ref{hpe}) one finds
\beq
[\hp_2,\phi(x)]=-i\pi(x)\hsp
[\hp_2,\pi(x)]=i\left(-\partial_x^2+V^{(2)}(\sl f(x,t))\right)\phi(x).
\eeq
Using the $\g$ equation of motion (\ref{geq}) the latter can be simplified
\beq
[\hp_2,\pi(x)]=-i\left( \ddot\g_B(x,0)\phi_0+\ddot\g_T(x,0)\phi_T+\ppin{k}\left(\ddot\g_k(x,0)\Bd{}+\ddot\g^*_k(x,0)\frac{B_{k}}{2\ok{}}\right)
\right).
\eeq
Iterating, it is not hard to see that
\bea
e^{i\hp_2 t}\phi(x) e^{-i\hp_2 t}&=&t\g_B(x,t)\pi_0+\g_B(x,t)\phi_0+\g_T(x,t)\phi_T+t\g_V(x,t)\pi_0\\
&&+\ppin{k}\left(\g_k(x,t)\Bd{}+\g^*_k(x,t)\frac{B_{k}}{2\ok{}}\right)\nonumber\\
e^{i\hp_2 t}\pi(x) e^{-i\hp_2 t}&=&\left(\g_B(x,t)+t\dot\g_B(x,t)\right)\pi_0+\dot\g_B(x,t)\phi_0+\dot\g_T(x,t)\phi_T+t\g_V(x,t)\pi_0\nonumber\\
&&+\ppin{k}\left(\dot\g_k(x,t)\Bd{}+\dot\g^*_k(x,t)\frac{B_{k}}{2\ok{}}\right).\nonumber
\eea
Or equivalently one can introduce a family of decompositions
\bea
\phi(x)&=&t\g_B(x,-t)\pi^{(t)}_0+\g_B(x,-t)\phi^{(t)}_0+\g_T(x,-t)\phi^{(t)}_T+t\g_V(x,-t)\pi^{(t)}_0\label{dect}\\
&&+\ppin{k}\left(\g_k(x,-t)\Btd{}+\g^*_k(x,-t)\frac{\Bt{}}{2\ok{}}\right)\nonumber\\
\pi(x)&=&\left(\g_B(x,-t)+t\dot\g_B(x,-t)\right)\pi^{(t)}_0+\dot\g_B(x,-t)\phi^{(t)}_0+\dot\g_T(x,-t)\phi^{(t)}_T+t\g_V(x,-t)\pi^{(t)}_0\nonumber\\
&&+\ppin{k}\left(\dot\g_k(x,-t)\Btd{}+\dot\g^*_k(x,-t)\frac{\Bt{}}{2\ok{}}\right)\nonumber
\eea
parametrized by $t$, where for any operator $\co$
\beq
\co^{(t)}=e^{i\hp_2 t}\co e^{-i\hp_2 t}. \label{cot}
\eeq
Note that Eq.~(\ref{cot}) implies that the algebra satisfied by these operators is independent of $t$
\beq
\left[\co^{(t_1)}_1,\co^{(t_1)}_2\right]=\
\left[\co^{(t_2)}_1,\co^{(t_2)}_2\right]=\
\left[\co_1,\co_2\right].
\eeq

Now clearly acting on a state with the operator $\co$ at time $t=0$ is equivalent to acting on it with $\co^{(t)}$ at time $t$, up to corrections of order $O(\sl)$.  Therefore we may use the algebra of the operators $\co$, that is to say the operators $\pi_0,\ \phi_0,\ \phi_t,\ B^\ddag_k$\ and $B_k$, to construct our comoving frame, one-soliton sector states at time $t=0$.  For example, we may impose that a state of interest is annihilated by some $\co_1$ and the excited state is created by acting with $\co_2$.   Then, at time $t$, our state will be annihilated by $\co^{(t)}_1$ and the excited state will be created by acting with $\co^{(t)}_2$.  This much is obvious, but then we may use the invertible expansions (\ref{dect}) to go between the $\co^{(t)}$ operators and the fields $\phi(x)$ and $\pi(x)$ at any time $t$, which allows us to use perturbation theory to study the static and dynamic properties of these states at any time.

\section{Remarks}

If the solution $f(x,t)$ is periodic then the parameter $t$ labeling the comoving frames is also periodic.  Therefore, after one period, one returns to the original frame.  If the original state was defined in the comoving frame by being annihilated by some set of operators $\co_i$, then after one period it will still be annihilated by those operators and so will still be in the same state.  In particular, it will not have decayed.  Thus, if such a state is sensible, for example if it does not posess classically unstable modes which would lead its perturbative expansion to be ill-defined, then it is stable at order $O(\lambda^0)$, in other words over timescales of order $O(1/m)$.  In this case one would have disproved the claim that oscillons decay already at the linear level in Refs.~\cite{hertzberg,tanmay}.  Indeed, this suggests more generally that the classical stability of a periodic classical solution generically implies an absence of decays via the emission of pairs of mesons.

So far we have only worked out one rather trivial example, the moving kink.  However, with this formalism developed, we hope to turn to more interesting applications.  We intend to approach quantum oscillons, estimating their lifetime, and also Rindler space, to see how quantum corrections affect the incoming radiation, trying to understand the remarkable yet perplexing results of Refs.~\cite{akh15,akh21,akh22,akh24}.

\section* {Acknowledgement}

\noindent
JE is supported by NSFC MianShang grants 11875296 and 11675223. KO is an awardee of the ANSO scholarship for UCAS Masters program.


\begin{thebibliography}{99}

\bibitem{sk}
H.~Segur and M.~D.~Kruskal,
``Nonexistence of Small Amplitude Breather Solutions in $\phi^4$ Theory,''
Phys. Rev. Lett. \textbf{58} (1987), 747-750
doi:10.1103/PhysRevLett.58.747

\bibitem{osc1}
E.~J.~Copeland, M.~Gleiser and H.~R.~Muller,
``Oscillons: Resonant configurations during bubble collapse,''
Phys. Rev. D \textbf{52} (1995), 1920-1933
doi:10.1103/PhysRevD.52.1920
[arXiv:hep-ph/9503217 [hep-ph]].

\bibitem{osc2}
M.~A.~Amin, R.~Easther, H.~Finkel, R.~Flauger and M.~P.~Hertzberg,
``Oscillons After Inflation,''
Phys. Rev. Lett. \textbf{108} (2012), 241302
doi:10.1103/PhysRevLett.108.241302
[arXiv:1106.3335 [astro-ph.CO]].

\bibitem{osc3}
P.~Dorey, A.~Gorina, T.~Roma\'nczukiewicz and Y.~Shnir,
``Collisions of weakly-bound kinks in the Christ-Lee model,''
[arXiv:2304.11710 [hep-th]].

\bibitem{hertzberg}
M.~P.~Hertzberg,
``Quantum Radiation of Oscillons,''
Phys. Rev. D \textbf{82} (2010), 045022
doi:10.1103/PhysRevD.82.045022
[arXiv:1003.3459 [hep-th]].

\bibitem{tanmay}
J.~Oll\'e, O.~Pujolas, T.~Vachaspati and G.~Zahariade,
``Quantum Evaporation of Classical Breathers,''
Phys. Rev. D \textbf{100} (2019) no.4, 045011
doi:10.1103/PhysRevD.100.045011
[arXiv:1904.12962 [hep-th]].



\bibitem{noiosc}
J.~Evslin, T.~Roma\'nczukiewicz and A.~Wereszczy\'nski,
``Quantum oscillons may be long-lived,''
JHEP \textbf{08} (2023), 182
doi:10.1007/JHEP08(2023)182
[arXiv:2305.18056 [hep-th]].

\bibitem{sug}
T.~Sugiyama,
``KINK - ANTIKINK COLLISIONS IN THE TWO-DIMENSIONAL PHI**4 MODEL,''
Prog. Theor. Phys. \textbf{61} (1979), 1550-1563
doi:10.1143/PTP.61.1550

\bibitem{csw} 
D.~K. Campbell, J.~F. Schonfeld and C.~A. Wingate,
``Resonance structure in kink-antikink interactions in $\phi^4$ theory,"
Physica \textbf{D9} (1983) 1.

\bibitem{frac91}
P.~Anninos, S.~Oliveira and R.~A.~Matzner,
``Fractal structure in the scalar lambda (phi**2-1)**2 theory,''
Phys. Rev. D \textbf{44} (1991), 1147-1160
doi:10.1103/PhysRevD.44.1147

\bibitem{pert0}
A.~Alonso Izquierdo, L.~M.~Nieto and J.~Queiroga-Nunes,
``Scattering between wobbling kinks,''
Phys. Rev. D \textbf{103} (2021) no.4, 045003
doi:10.1103/PhysRevD.103.045003
[arXiv:2007.15517 [hep-th]].

\bibitem{pert1}
C.~Adam, P.~Dorey, A.~Garcia Martin-Caro, M.~Huidobro, K.~Oles, T.~Romanczukiewicz, Y.~Shnir and A.~Wereszczynski,
``Multikink scattering in the \ensuremath{\phi}6 model revisited,''
Phys. Rev. D \textbf{106} (2022) no.12, 125003
doi:10.1103/PhysRevD.106.125003
[arXiv:2209.08849 [hep-th]].

\bibitem{pert2}
S.~Navarro-Obreg\'on, L.~M.~Nieto and J.~M.~Queiruga,
``Inclusion of radiation in the collective coordinate method approach of the \ensuremath{\phi}4 model,''
Phys. Rev. E \textbf{108} (2023) no.4, 044216
doi:10.1103/PhysRevE.108.044216
[arXiv:2305.00497 [hep-th]].

\bibitem{pert3}
A.~M.~Marjaneh, A.~Ghaani and K.~Javidan,
``Kinks scattering in deformed $\varphi^6$ model,''
[arXiv:2309.12599 [nlin.PS]].

\bibitem{qpert23}
M.~Mukhopadhyay and T.~Vachaspati,
``Resonance structures in kink-antikink scattering in a quantum vacuum,''
Phys. Rev. D \textbf{107} (2023) no.11, 116017
doi:10.1103/PhysRevD.107.116017
[arXiv:2303.03415 [hep-th]].

\bibitem{akh21}
E.~T.~Akhmedov, K.~V.~Bazarov and D.~V.~Diakonov,
``Quantum fields in the future Rindler wedge,''
Phys. Rev. D \textbf{104} (2021) no.8, 085008
doi:10.1103/PhysRevD.104.085008
[arXiv:2106.01791 [hep-th]].



\bibitem{gw22}
N.~Graham and H.~Weigel,
``Quantum corrections to soliton energies,''
Int. J. Mod. Phys. A \textbf{37} (2022) no.19, 2241004
doi:10.1142/S0217751X22410044
[arXiv:2201.12131 [hep-th]].

\bibitem{cq1}
T.~Vachaspati and G.~Zahariade,
``Classical-quantum correspondence and backreaction,''
Phys. Rev. D \textbf{98} (2018) no.6, 065002
doi:10.1103/PhysRevD.98.065002
[arXiv:1806.05196 [hep-th]].

\bibitem{cq2}
T.~Vachaspati and G.~Zahariade,
``Classical-Quantum Correspondence for Fields,''
JCAP \textbf{09} (2019), 015
doi:10.1088/1475-7516/2019/09/015
[arXiv:1807.10282 [hep-th]].

\bibitem{gs74}
J.~L.~Gervais and B.~Sakita,
``Extended Particles in Quantum Field Theories,''
Phys. Rev. D \textbf{11} (1975), 2943
doi:10.1103/PhysRevD.11.2943

\bibitem{gj76}
J.~L.~Gervais and A.~Jevicki,
``Point Canonical Transformations in Path Integral,''
Nucl. Phys. B \textbf{110} (1976), 93-112
doi:10.1016/0550-3213(76)90422-3


\bibitem{mekink}
J.~Evslin,
``Manifestly Finite Derivation of the Quantum Kink Mass,''
JHEP \textbf{11} (2019), 161
doi:10.1007/JHEP11(2019)161
[arXiv:1908.06710 [hep-th]].

\bibitem{me2loop}
J.~Evslin and H.~Guo,
``Two-Loop Scalar Kinks,''
Phys. Rev. D \textbf{103} (2021) no.12, 125011
doi:10.1103/PhysRevD.103.125011
[arXiv:2012.04912 [hep-th]].

\bibitem{cahill76}
K.~E.~Cahill, A.~Comtet and R.~J.~Glauber,
``Mass Formulas for Static Solitons,''
Phys. Lett. B \textbf{64} (1976), 283-285
doi:10.1016/0370-2693(76)90202-1

\bibitem{rebhan}
  A.~Rebhan and P.~van Nieuwenhuizen,
  ``No saturation of the quantum Bogomolnyi bound by two-dimensional supersymmetric solitons,''
  Nucl.\ Phys.\ B {\bf 508} (1997) 449
  doi:10.1016/S0550-3213(97)00625-1, 10.1016/S0550-3213(97)80021-1
 [hep-th/9707163].

\bibitem{taylor78}
J.~G.~Taylor,
``Solitons as Infinite Constituent Bound States,''
Annals Phys. \textbf{115} (1978), 153
doi:10.1016/0003-4916(78)90179-3

\bibitem{cocorr23}
L.~Berezhiani, G.~Cintia and M.~Zantedeschi,
``Perturbative Construction of Coherent States,''
[arXiv:2311.18650 [hep-th]].

\bibitem{memult}
H.~Liu, J.~Evslin and B.~Zhang,
``Meson production from kink-meson scattering,''
Phys. Rev. D \textbf{107} (2023) no.2, 025012
doi:10.1103/PhysRevD.107.025012
[arXiv:2211.01794 [hep-th]].

\bibitem{mestokes}
J.~Evslin and H.~Liu,
``(Anti-)Stokes scattering on kinks,''
JHEP \textbf{03} (2023), 095
doi:10.1007/JHEP03(2023)095
[arXiv:2301.04099 [hep-th]].

\bibitem{elastico}
J.~Evslin and H.~Liu,
``Elastic Kink-Meson Scattering,''
[arXiv:2311.14369 [hep-th]].

\bibitem{memove}
J.~Evslin,
``Moving kinks and their wave packets,''
Phys. Rev. D \textbf{105} (2022) no.10, 105001
doi:10.1103/PhysRevD.105.105001
[arXiv:2202.04905 [hep-th]].

\bibitem{meff}
J.~Evslin,
``Kink form factors,''
JHEP \textbf{07} (2022), 033
doi:10.1007/JHEP07(2022)033
[arXiv:2203.15445 [hep-th]].

\bibitem{hengyuan}
H.~Guo,
``Leading quantum correction to the \ensuremath{\Phi}4 kink form factor,''
Phys. Rev. D \textbf{106} (2022) no.9, 096001
doi:10.1103/PhysRevD.106.096001
[arXiv:2209.03650 [hep-th]].

\bibitem{wstabile}
H.~Weigel,
``Quantum Instabilities of Solitons,''
AIP Conf. Proc. \textbf{2116} (2019) no.1, 170002
doi:10.1063/1.5114153
[arXiv:1907.10942 [hep-th]].

\bibitem{akh15}
E.~T.~Akhmedov, H.~Godazgar and F.~K.~Popov,
``Hawking radiation and secularly growing loop corrections,''
Phys. Rev. D \textbf{93} (2016) no.2, 024029
doi:10.1103/PhysRevD.93.024029
[arXiv:1508.07500 [hep-th]].

\bibitem{akh22}
E.~T.~Akhmedov and P.~A.~Anempodistov,
``Loop corrections to cosmological particle creation,''
Phys. Rev. D \textbf{105} (2022) no.10, 105019
doi:10.1103/PhysRevD.105.105019
[arXiv:2204.01388 [hep-th]].

\bibitem{akh24}
E.~T.~Akhmedov, P.~A.~Anempodistov and K.~V.~Bazarov,
``On a non trivial self-consistent backreaction of quantum fields in 2D dilaton gravity,''
[arXiv:2401.07645 [hep-th]].

\end{thebibliography}

\begin{thebibliography}{99}

\bibitem{weigel89}
B.~Schwesinger, H.~Weigel, G.~Holzwarth and A.~Hayashi,
``The Skyrme Soliton in Pion, Vector and Scalar Meson Fields: $\pi N$ Scattering and Photoproduction,''
Phys. Rept. \textbf{173} (1989), 173
doi:10.1016/0370-1573(89)90022-7

\bibitem{gs74}
J.~L.~Gervais and B.~Sakita,
``Extended Particles in Quantum Field Theories,''
Phys. Rev. D \textbf{11} (1975), 2943
doi:10.1103/PhysRevD.11.2943

\bibitem{tom75}
E.~Tomboulis,
``Canonical Quantization of Nonlinear Waves,''
Phys. Rev. D \textbf{12} (1975), 1678
doi:10.1103/PhysRevD.12.1678

\bibitem{uehara91}
M.~Uehara, A.~Hayashi and S.~Saito,
``Meson - soliton scattering with full recoil in standard collective coordinate quantization,''
Nucl. Phys. A \textbf{534} (1991), 680-696
doi:10.1016/0375-9474(91)90466-J

\bibitem{hayashi92}
A.~Hayashi, S.~Saito and M.~Uehara,
``New formulation of pion - nucleon scattering and soft pion theorems in the Skyrme model,''
Phys. Rev. D \textbf{46} (1992), 4856-4867
doi:10.1103/PhysRevD.46.4856

\bibitem{erase}
M.~Bachmaier, G.~Dvali and J.~S.~Valbuena-Berm\'udez,
``Radiation emission during the erasure of magnetic monopoles,''
Phys. Rev. D \textbf{108} (2023) no.10, 103501
doi:10.1103/PhysRevD.108.103501
[arXiv:2306.12958 [hep-th]].

\bibitem{adiabatic84}
A.~Hayashi, G.~Eckart, G.~Holzwarth and H.~Walliser,
``Pion Nucleon Scattering Phase Shifts in the Skyrme Model,''
Phys. Lett. B \textbf{147} (1984), 5-9
doi:10.1016/0370-2693(84)90581-1

\bibitem{diakonov97}
D.~Diakonov, V.~Petrov and M.~V.~Polyakov,
``Exotic anti-decuplet of baryons: Prediction from chiral solitons,''
Z. Phys. A \textbf{359} (1997), 305-314
doi:10.1007/s002180050406
[arXiv:hep-ph/9703373 [hep-ph]].

\bibitem{ellis04}
J.~R.~Ellis, M.~Karliner and M.~Praszalowicz,
``Chiral soliton predictions for exotic baryons,''
JHEP \textbf{05} (2004), 002
doi:10.1088/1126-6708/2004/05/002
[arXiv:hep-ph/0401127 [hep-ph]].

\bibitem{nakano03}
T.~Nakano \textit{et al.} [LEPS],
``Evidence for a narrow S = +1 baryon resonance in photoproduction from the neutron,''
Phys. Rev. Lett. \textbf{91} (2003), 012002
doi:10.1103/PhysRevLett.91.012002
[arXiv:hep-ex/0301020 [hep-ex]].

\bibitem{notheta}
T.~Liu, Y.~Mao and B.~Q.~Ma,
``Present status on experimental search for pentaquarks,''
Int. J. Mod. Phys. A \textbf{29} (2014) no.13, 1430020
doi:10.1142/S0217751X14300208
[arXiv:1403.4455 [hep-ex]].

\bibitem{petrov16}
V.~Petrov,
``Soliton Model for Baryons,''
Acta Phys. Polon. B \textbf{47} (2016), 59-95
doi:10.5506/APhysPolB.47.59

\bibitem{weigel18}
H.~Weigel,
``Exotic Baryons in Chiral Soliton Models,''
Universe \textbf{4} (2018) no.12, 142
doi:10.3390/universe4120142
[arXiv:1811.12128 [hep-ph]].

\bibitem{chris23}
S.~B.~Gudnason and C.~Halcrow,
``Quantum binding energies in the Skyrme model,''
[arXiv:2307.09272 [hep-th]].

\bibitem{bjarke23}
S.~B.~Gudnason,
``Nonlinear rigid-body quantization of Skyrmions,''
[arXiv:2311.11667 [hep-th]].

\bibitem{mekink}
J.~Evslin,
``Manifestly Finite Derivation of the Quantum Kink Mass,''
JHEP \textbf{11} (2019), 161
doi:10.1007/JHEP11(2019)161
[arXiv:1908.06710 [hep-th]].

\bibitem{me2loop}
J.~Evslin and H.~Guo,
``Two-Loop Scalar Kinks,''
Phys. Rev. D \textbf{103} (2021) no.12, 125011
doi:10.1103/PhysRevD.103.125011
[arXiv:2012.04912 [hep-th]].

\bibitem{wstabile}
H.~Weigel,
``Quantum Instabilities of Solitons,''
AIP Conf. Proc. \textbf{2116} (2019) no.1, 170002
doi:10.1063/1.5114153
[arXiv:1907.10942 [hep-th]].


\bibitem{cahill76}
K.~E.~Cahill, A.~Comtet and R.~J.~Glauber,
``Mass Formulas for Static Solitons,''
Phys. Lett. B \textbf{64} (1976), 283-285
doi:10.1016/0370-2693(76)90202-1

\bibitem{menorm}
J.~Evslin and H.~Liu,
``A reduced inner product for kink states,''
JHEP \textbf{03} (2023), 070
doi:10.1007/JHEP03(2023)070
[arXiv:2212.10344 [hep-th]].

\bibitem{ellong}
J.~Evslin and H. Liu,
In preparation.

\bibitem{metad}
J.~Evslin and H.~Guo,
``Removing tadpoles in a soliton sector,''
JHEP \textbf{11} (2021), 128
doi:10.1007/JHEP11(2021)128
[arXiv:2110.00234 [hep-th]].

\bibitem{alberto}
J.~Evslin and A.~Garc\'\i{}a Mart\'\i{}n-Caro,
``Spontaneous emission from excited quantum kinks,''
JHEP \textbf{12} (2022), 111
doi:10.1007/JHEP12(2022)111
[arXiv:2210.13791 [hep-th]].

\bibitem{memult}
H.~Liu, J.~Evslin and B.~Zhang,
``Meson production from kink-meson scattering,''
Phys. Rev. D \textbf{107} (2023) no.2, 025012
doi:10.1103/PhysRevD.107.025012
[arXiv:2211.01794 [hep-th]].



\bibitem{mech22}
F.~Blaschke and O.~N.~Karpisek,
``Mechanization of scalar field theory in 1+1 dimensions,''
PTEP \textbf{2022} (2022) no.10, 103A01
doi:10.1093/ptep/ptac104
[arXiv:2202.05675 [hep-th]].

\bibitem{multi22}
C.~Adam, P.~Dorey, A.~Garcia Martin-Caro, M.~Huidobro, K.~Oles, T.~Romanczukiewicz, Y.~Shnir and A.~Wereszczynski,
``Multikink scattering in the \ensuremath{\phi}6 model revisited,''
Phys. Rev. D \textbf{106} (2022) no.12, 125003
doi:10.1103/PhysRevD.106.125003
[arXiv:2209.08849 [hep-th]].

\bibitem{dorey23}
P.~Dorey, A.~Gorina, T.~Roma\'nczukiewicz and Y.~Shnir,
``Collisions of weakly-bound kinks in the Christ-Lee model,''
JHEP \textbf{09} (2023), 045
doi:10.1007/JHEP09(2023)045
[arXiv:2304.11710 [hep-th]].

\bibitem{nav23}
S.~Navarro-Obreg\'on, L.~M.~Nieto and J.~M.~Queiruga,
``Inclusion of radiation in the CCM approach of the $\phi^4$ model,''
[arXiv:2305.00497 [hep-th]].

\bibitem{hahne23}
F.~M.~Hahne and P.~Klimas,
``Scattering of compact kinks,''
[arXiv:2311.09494 [hep-th]].

\bibitem{doreyprl}
P.~Dorey, K.~Mersh, T.~Romanczukiewicz and Y.~Shnir,
``Kink-antikink collisions in the $\phi^6$ model,''
Phys. Rev. Lett. \textbf{107} (2011), 091602
doi:10.1103/PhysRevLett.107.091602
[arXiv:1101.5951 [hep-th]].

\bibitem{kinkquantscat}
M.~Mukhopadhyay and T.~Vachaspati,
``Resonance structures in kink-antikink scattering in a quantum vacuum,''
Phys. Rev. D \textbf{107} (2023) no.11, 116017
doi:10.1103/PhysRevD.107.116017
[arXiv:2303.03415 [hep-th]].

\bibitem{alonso14}
A.~Alonso-Izquierdo and J.~M.~Guilarte,
``Quantum-induced interactions in the moduli space of degenerate BPS domain walls,''
JHEP \textbf{01} (2014), 125
doi:10.1007/JHEP01(2014)125
[arXiv:1307.0740 [hep-th]].

\bibitem{vach23}
O.~Albayrak and T.~Vachaspati,
``Creating kinks with quantum mediation,''
[arXiv:2308.01962 [hep-ph]].

\end{thebibliography}

\begin{thebibliography}{99}

\bibitem{gjscc}
J.~L.~Gervais, A.~Jevicki and B.~Sakita,
``Collective Coordinate Method for Quantization of Extended Systems,''
Phys. Rept. \textbf{23} (1976), 281-293
doi:10.1016/0370-1573(76)90049-1

\bibitem{dhn2}
  R.~F.~Dashen, B.~Hasslacher and A.~Neveu,
  ``Nonperturbative Methods and Extended Hadron Models in Field Theory 2. Two-Dimensional Models and Extended Hadrons,''
  Phys.\ Rev.\ D {\bf 10} (1974) 4130.
 doi:10.1103/PhysRevD.10.4130

\bibitem{wrev}
N.~Graham and H.~Weigel,
``Quantum corrections to soliton energies,''
Int. J. Mod. Phys. A \textbf{37} (2022) no.19, 2241004
doi:10.1142/S0217751X22410044
[arXiv:2201.12131 [hep-th]].

\bibitem{gj76}
J.~L.~Gervais and A.~Jevicki,
``Point Canonical Transformations in Path Integral,''
Nucl. Phys. B \textbf{110} (1976), 93-112
doi:10.1016/0550-3213(76)90422-3

\bibitem{vega}
H.~J.~de Vega,
``Two-Loop Quantum Corrections to the Soliton Mass in Two-Dimensional Scalar Field Theories,''
Nucl. Phys. B \textbf{115} (1976), 411-428
doi:10.1016/0550-3213(76)90497-1

\bibitem{verwaest}
J.~Verwaest,
``Higher Order Correction to the Sine-Gordon Soliton Mass,''
Nucl. Phys. B \textbf{123} (1977), 100-108
doi:10.1016/0550-3213(77)90343-1

\bibitem{shifmananomalia}
M.~A.~Shifman, A.~I.~Vainshtein and M.~B.~Voloshin,
``Anomaly and quantum corrections to solitons in two-dimensional theories with minimal supersymmetry,''
Phys. Rev. D \textbf{59} (1999), 045016
doi:10.1103/PhysRevD.59.045016
[arXiv:hep-th/9810068 [hep-th]].

\bibitem{hayashi1}
A.~Hayashi, S.~Saito and M.~Uehara,
``Pion - nucleon scattering in the Skyrme model and the P wave Born amplitudes,''
Phys. Rev. D \textbf{43} (1991), 1520-1531
doi:10.1103/PhysRevD.43.1520

\bibitem{hayashi2}
A.~Hayashi, S.~Saito and M.~Uehara,
``Pion - nucleon scattering in the soliton model,''
Prog. Theor. Phys. Suppl. \textbf{109} (1992), 45-72
doi:10.1143/PTPS.109.45


\bibitem{accel}
I.~V.~Melnikov, C.~Papageorgakis and A.~B.~Royston,
``Accelerating solitons,''
Phys. Rev. D \textbf{102} (2020) no.12, 125002
doi:10.1103/PhysRevD.102.125002
[arXiv:2007.11028 [hep-th]].

\bibitem{andyprl}
I.~V.~Melnikov, C.~Papageorgakis and A.~B.~Royston,
``Forced Soliton Equation and Semiclassical Soliton Form Factors,''
Phys. Rev. Lett. \textbf{125} (2020) no.23, 231601
doi:10.1103/PhysRevLett.125.231601
[arXiv:2010.10381 [hep-th]].

\bibitem{raggio}
J.~F.~Wheater and P.~D.~Xavier,
``The Size of a Soliton,''
[arXiv:2207.01274 [hep-th]].

\bibitem{mekink}
J.~Evslin,
``Manifestly Finite Derivation of the Quantum Kink Mass,''
JHEP \textbf{11} (2019), 161
doi:10.1007/JHEP11(2019)161
[arXiv:1908.06710 [hep-th]].

\bibitem{me2loop}
J.~Evslin and H.~Guo,
``Two-Loop Scalar Kinks,''
Phys. Rev. D \textbf{103} (2021) no.12, 125011
doi:10.1103/PhysRevD.103.125011
[arXiv:2012.04912 [hep-th]].


\bibitem{memono}
J.~Evslin and S.~B.~Gudnason,
``Dwarf Galaxy Sized Monopoles as Dark Matter?,''
[arXiv:1202.0560 [astro-ph.CO]].


\bibitem{fuzzy14}
H.~Y.~Schive, T.~Chiueh and T.~Broadhurst,
``Cosmic Structure as the Quantum Interference of a Coherent Dark Wave,''
Nature Phys. \textbf{10} (2014), 496-499
doi:10.1038/nphys2996
[arXiv:1406.6586 [astro-ph.GA]].

\bibitem{impure}
C.~Adam, K.~Oles, J.~M.~Queiruga, T.~Romanczukiewicz and A.~Wereszczynski,
``Solvable self-dual impurity models,''
JHEP \textbf{07} (2019), 150
doi:10.1007/JHEP07(2019)150
[arXiv:1905.06080 [hep-th]].

\bibitem{chris}
J.~Evslin, C.~Halcrow, T.~Romanczukiewicz and A.~Wereszczynski,
``Spectral walls at one loop,''
Phys. Rev. D \textbf{105} (2022) no.12, 125002
doi:10.1103/PhysRevD.105.125002
[arXiv:2202.08249 [hep-th]].

\bibitem{hayashirep}
B.~Schwesinger, H.~Weigel, G.~Holzwarth and A.~Hayashi,
``The Skyrme Soliton in Pion, Vector and Scalar Meson Fields: $\pi N$ Scattering and Photoproduction,''
Phys. Rept. \textbf{173} (1989), 173
doi:10.1016/0370-1573(89)90022-7

\bibitem{skyrme}
T.~H.~R.~Skyrme,
``A Nonlinear field theory,''
Proc. Roy. Soc. Lond. A \textbf{260} (1961), 127-138
doi:10.1098/rspa.1961.0018

\bibitem{smorg}
S.~B.~Gudnason and C.~Halcrow,
``A Sm\"orgasbord of Skyrmions,''
JHEP \textbf{08} (2022), 117
doi:10.1007/JHEP08(2022)117
[arXiv:2202.01792 [hep-th]].

\bibitem{point22}
M.~A.~A.~Martin, R.~Schlesier and J.~Zahn,
``The semiclassical energy density of kinks and solitons,''
[arXiv:2204.08785 [hep-th]].

\bibitem{phi42loop}
J.~Evslin,
``\ensuremath{\phi^4} kink mass at two loops,''
Phys. Rev. D \textbf{104} (2021) no.8, 085013
doi:10.1103/PhysRevD.104.085013
[arXiv:2104.07991 [hep-th]].

\bibitem{menormal}
J.~Evslin and H.~Guo,
``Excited Kinks as Quantum States,''
Eur. Phys. J. C \textbf{81} (2021) no.10, 936
doi:10.1140/epjc/s10052-021-09739-9
[arXiv:2104.03612 [hep-th]].

\bibitem{hengyuan}
H.~Guo,
``Leading quantum correction to the \ensuremath{\Phi}4 kink form factor,''
Phys. Rev. D \textbf{106} (2022) no.9, 096001
doi:10.1103/PhysRevD.106.096001
[arXiv:2209.03650 [hep-th]].

\bibitem{alberto}
J.~Evslin and A.~Garc\'\i{}a Mart\'\i{}n-Caro,
``Spontaneous emission from excited quantum kinks,''
JHEP \textbf{12} (2022), 111
doi:10.1007/JHEP12(2022)111
[arXiv:2210.13791 [hep-th]].

\bibitem{memult}
H.~Liu, J.~Evslin and B.~Zhang,
``Meson production from kink-meson scattering,''
Phys. Rev. D \textbf{107} (2023) no.2, 025012
doi:10.1103/PhysRevD.107.025012
[arXiv:2211.01794 [hep-th]].


\bibitem{wstabile}
H.~Weigel,
``Quantum Instabilities of Solitons,''
AIP Conf. Proc. \textbf{2116} (2019) no.1, 170002
doi:10.1063/1.5114153
[arXiv:1907.10942 [hep-th]].

\bibitem{cahill76}
K.~E.~Cahill, A.~Comtet and R.~J.~Glauber,
``Mass Formulas for Static Solitons,''
Phys. Lett. B \textbf{64} (1976), 283-285
doi:10.1016/0370-2693(76)90202-1

\bibitem{mewick}
J.~Evslin,
``Normal ordering normal modes,''
Eur. Phys. J. C \textbf{81} (2021) no.1, 92
doi:10.1140/epjc/s10052-021-08890-7
[arXiv:2007.05741 [hep-th]].

\bibitem{shapeinter}
A.~Alonso-Izquierdo, D.~Migu\'elez-Caballero, L.~M.~Nieto and J.~Queiroga-Nunes,
``Wobbling kinks in a two-component scalar field theory: Interaction between shape modes,''
[arXiv:2207.10989 [hep-th]].

\bibitem{wshifman}
H.~Weigel and N.~Graham,
``Vacuum polarization energy of the Shifman\textendash{}Voloshin soliton,''
Phys. Lett. B \textbf{783} (2018), 434-439
doi:10.1016/j.physletb.2018.07.027
[arXiv:1806.07584 [hep-th]].

\bibitem{takyi1}
I.~Takyi, M.~K.~Matfunjwa and H.~Weigel,
``Quantum corrections to solitons in the $\Phi^8$ model,''
Phys. Rev. D \textbf{102} (2020) no.11, 116004
doi:10.1103/PhysRevD.102.116004
[arXiv:2010.07182 [hep-th]].

\bibitem{takyi2}
I.~Takyi, B.~Barnes and J.~Ackora-Prah,
``Vacuum Polarization Energy of the Kinks in the Sinh-Deformed Models,''
Turk. J. Phys. \textbf{45} (2021), 194-206
[arXiv:2012.12343 [hep-th]].

\bibitem{yuan1}
Y.~Zhong,
``Normal modes for two-dimensional gravitating kinks,''
Phys. Lett. B \textbf{827} (2022), 136947
doi:10.1016/j.physletb.2022.136947
[arXiv:2112.08683 [hep-th]].

\bibitem{yuan2}
Y.~Zhong,
``Singular P\"oschl-Teller II potentials and gravitating kinks,''
JHEP \textbf{09} (2022), 165
doi:10.1007/JHEP09(2022)165
[arXiv:2207.12681 [hep-th]].

\bibitem{quantosc}
M.~P.~Hertzberg,
``Quantum Radiation of Oscillons,''
Phys. Rev. D \textbf{82} (2010), 045022
doi:10.1103/PhysRevD.82.045022
[arXiv:1003.3459 [hep-th]].

\bibitem{kovbreather}
A.~Kovtun,
``Analytical computation of quantum corrections to a nontopological soliton within the saddle-point approximation,''
Phys. Rev. D \textbf{105} (2022) no.3, 036011
doi:10.1103/PhysRevD.105.036011
[arXiv:2110.05222 [hep-th]].

\end{thebibliography}

\begin{thebibliography}{99}

\bibitem{csw} D.~K. Campbell, J.~F. Schonfeld and C.~A. Wingate,
``Resonance structure in kink-antikink interactions in $\phi^4$ theory,"
Physica \textbf{D9} (1983) 1.

\bibitem{osc}
H.~Segur and M.~D.~Kruskal,
``Nonexistence of Small Amplitude Breather Solutions in $\phi^4$ Theory,''
Phys. Rev. Lett. \textbf{58} (1987), 747-750
doi:10.1103/PhysRevLett.58.747

\bibitem{osc3d}
G.~Fodor, P.~Forgacs, P.~Grandclement and I.~Racz,
``Oscillons and Quasi-breathers in the phi**4 Klein-Gordon model,''
Phys. Rev. D \textbf{74} (2006), 124003
doi:10.1103/PhysRevD.74.124003
[arXiv:hep-th/0609023 [hep-th]].

\bibitem{doreyf6}
P.~Dorey, K.~Mersh, T.~Romanczukiewicz and Y.~Shnir,
``Kink-antikink collisions in the $\phi^6$ model,''
Phys. Rev. Lett. \textbf{107} (2011), 091602
doi:10.1103/PhysRevLett.107.091602
[arXiv:1101.5951 [hep-th]].

\bibitem{multex22a}
F.~C.~Simas, K.~Z.~Nobrega, D.~Bazeia and A.~R.~Gomes,
``Degeneracy and kink scattering in a two coupled scalar field model in $(1,1)$ dimensions,''
[arXiv:2201.03372 [hep-th]].

\bibitem{multex22b}
A.~Moradi Marjaneh, F.~C.~Simas and D.~Bazeia,
``Collisions of kinks in deformed  \ensuremath{\varphi}${}^4$ and \ensuremath{\varphi}${}^6$ models,''
Chaos Solitons and Fractals: the interdisciplinary journal of Nonlinear Science and Nonequilibrium and Complex Phenomena \textbf{164} (2022), 112723
doi:10.1016/j.chaos.2022.112723
[arXiv:2207.00835 [hep-th]].

\bibitem{sfal21}
C.~Adam, D.~Ciurla, K.~Oles, T.~Romanczukiewicz and A.~Wereszczynski,
``Sphalerons and resonance phenomenon in kink-antikink collisions,''
Phys. Rev. D \textbf{104} (2021) no.10, 105022
doi:10.1103/PhysRevD.104.105022
[arXiv:2109.01834 [hep-th]].

\bibitem{col22}
C.~Adam, P.~Dorey, A.~Garcia Martin-Caro, M.~Huidobro, K.~Oles, T.~Romanczukiewicz, Y.~Shnir and A.~Wereszczynski,
``Multikink scattering in the $\phi^6$ model revisited,''
[arXiv:2209.08849 [hep-th]].

\bibitem{mm}
N.~S.~Manton and H.~Merabet,
``$\phi^4$ kinks: Gradient flow and dynamics,''
Nonlinearity \textbf{10} (1997), 3
doi:10.1088/0951-7715/10/1/002
[arXiv:hep-th/9605038 [hep-th]].

\bibitem{tomrad1}
T.~Romanczukiewicz,
``Interaction between kink and radiation in phi**4 model,''
Acta Phys. Polon. B \textbf{35} (2004), 523-540
[arXiv:hep-th/0303058 [hep-th]].

\bibitem{tomrad2}
T.~Romanczukiewicz,
``Interaction between topological defects and radiation,''
Acta Phys. Polon. B \textbf{36} (2005), 3877-3887

\bibitem{tomrad3}
P.~Forgacs, A.~Lukacs and T.~Romanczukiewicz,
``Negative radiation pressure exerted on kinks,''
Phys. Rev. D \textbf{77} (2008), 125012
doi:10.1103/PhysRevD.77.125012
[arXiv:0802.0080 [hep-th]].

\bibitem{mekink}
J.~Evslin,
``Manifestly Finite Derivation of the Quantum Kink Mass,''
JHEP \textbf{11} (2019), 161
doi:10.1007/JHEP11(2019)161
[arXiv:1908.06710 [hep-th]].

\bibitem{me2loop}
J.~Evslin and H.~Guo,
``Two-Loop Scalar Kinks,''
Phys. Rev. D \textbf{103} (2021) no.12, 125011
doi:10.1103/PhysRevD.103.125011
[arXiv:2012.04912 [hep-th]].

\bibitem{gjscc}
J.~L.~Gervais, A.~Jevicki and B.~Sakita,
``Collective Coordinate Method for Quantization of Extended Systems,''
Phys. Rept. \textbf{23} (1976), 281-293
doi:10.1016/0370-1573(76)90049-1

\bibitem{gj76}
J.~L.~Gervais and A.~Jevicki,
``Point Canonical Transformations in Path Integral,''
Nucl. Phys. B \textbf{110} (1976), 93-112
doi:10.1016/0550-3213(76)90422-3

\bibitem{wstabile}
H.~Weigel,
``Quantum Instabilities of Solitons,''
AIP Conf. Proc. \textbf{2116} (2019) no.1, 170002
doi:10.1063/1.5114153
[arXiv:1907.10942 [hep-th]].

\bibitem{cahill76}
K.~E.~Cahill, A.~Comtet and R.~J.~Glauber,
``Mass Formulas for Static Solitons,''
Phys. Lett. B \textbf{64} (1976), 283-285
doi:10.1016/0370-2693(76)90202-1

\bibitem{phi42loop}
J.~Evslin,
``\ensuremath{\phi^4} kink mass at two loops,''
Phys. Rev. D \textbf{104} (2021) no.8, 085013
doi:10.1103/PhysRevD.104.085013
[arXiv:2104.07991 [hep-th]].

\bibitem{merif}
J.~Evslin and H.~Liu,
``Quantum Reflective Kinks,''
[arXiv:2210.12725 [hep-th]].

\end{thebibliography}
\end{document}